\documentstyle[epsf,12pt,cite]{article}

\begin{document}

\newcommand{\referee}[1]{#1}
\newcommand{\camera}[1]{\relax}

% -----------------------------------------------------------
% my personal definitions for math symbols, etc:

\renewcommand{\epsilon}{\varepsilon}
\newcommand{\refcite}[1]{\cite{#1}}

\newcommand{\figuno}{1}       %  fig1Largo(ro lejos)
\newcommand{\figdos}{2}       %  fig2Largo(ro medio)
\newcommand{\figtres}{3}      %  fig3Largo(ro cerca)
\newcommand{\figcuatro}{5}    %  fig4Largo(parac mejor)
\newcommand{\figcuatrob}{7}   %  fig4~bLargo(GGLnoE)
\newcommand{\figseis}{9}      %  fig6Largo(dif bk 10y25)
\newcommand{\figcinco}{10}    %  fig5Largo(dif Tc 10y25)
\newcommand{\figsiete}{6}     %  fig7Largo(Suzuki)
\newcommand{\figocho}{8}      %  fig8Largo(xic y Tc vs x)
\newcommand{\figTci}{4}       %  fig9Largo( Tci(x) )

\newcommand\eq[1]{(\ref{#1})}

\newcommand\eTci{\mbox{$\epsilon_{T_{CI}}$}}
\newcommand\Tsuper{\mbox{$T^C$}}
\newcommand\esuper{\mbox{$\epsilon^C$}}
\newcommand\epsilonsuper{\mbox{$\epsilon^C$}}
\newcommand\Tc{\mbox{$T_C$}}
\newcommand\Tci{\mbox{$T_{CI}$}}
\newcommand\Tcip{\mbox{$T_{CI}^+$}}
\newcommand\Tcim{\mbox{$T_{CI}^-$}}
\newcommand\Tco{\mbox{$T_{C0}$}}
\newcommand\DTci{\mbox{$\Delta T_{CI}$}}
\newcommand\TLB{\mbox{$T_{B}^L$}}
\newcommand\TBL{\mbox{$T_{B}^L$}}
\newcommand\Tpseudo{\mbox{$T^*$}}
\newcommand\LSCOf{La$_{2-x}$Sr$_x$CuO$_4$}
\newcommand\YBCOf{YBa$_2$Cu$_3$O$_x$}
\newcommand\BSCOf{Bi$_2$Sr$_2$CaCu$_2$O$_{8+x}$}
\newcommand\Taliof{Tl$_2$Ba$_2$Ca$_2$Cu$_3$O$_{10}$}
\newcommand\LSCO{LS$x$CO}
\newcommand\YBCO{YBCO$x$}
\newcommand\BSCO{BSCO}

\newcommand\gsim{\stackrel{>}{_\sim}}
\newcommand\lsim{\stackrel{<}{_\sim}}

\newcommand\xie{\mbox{$\xi(\epsilon)$}}
\newcommand\xio{\mbox{$\xi(0)$}}
\newcommand\xisubo{\mbox{$\xi_0$}}
\newcommand\xiabe{\mbox{$\xi_{ab}(\epsilon)$}}
\newcommand\xiab{\mbox{$\xi_{ab}$}}
\newcommand\xiabo{\mbox{$\xi_{ab}(0)$}}
\newcommand\xiabsubo{\mbox{$\xi_{ab0}$}}
\newcommand\xice{\mbox{$\xi_{c}(\epsilon)$}}
\newcommand\xico{\mbox{$\xi_{c}(0)$}}

\newcommand\BLD{\mbox{$B_{\rm LD}$}}
\newcommand\kB{\mbox{$k_{\rm B}$}}

\newcommand\rhoabT{\mbox{$\rho_{ab}(T)$}}
\newcommand\rhoab{\mbox{$\rho_{ab}$}}
\newcommand\rhoabB{\mbox{$\rho_{abB}$}}
\newcommand\rhoabBT{\mbox{$\rho_{abB}(T)$}}

\newcommand\roabT{\mbox{$\rho_{ab}(T)$}}
\newcommand\roab{\mbox{$\rho_{ab}$}}
\newcommand\roabB{\mbox{$\rho_{abB}$}}
\newcommand\roabBT{\mbox{$\rho_{abB}(T)$}}

\newcommand\sabT{\mbox{$\sigma_{ab}(T)$}}
\newcommand\sab{\mbox{$\sigma_{ab}$}}
\newcommand\DsabT{\mbox{$\Delta\sigma_{ab}(T)$}}
\newcommand\Dsabe{\mbox{$\Delta\sigma_{ab}(\epsilon)$}}
\newcommand\DsabeTci{\mbox{$\Delta\sigma_{ab}(\epsilon_{T_{CI}})$}}

\newcommand\Dsab{\mbox{$\Delta\sigma_{ab}$}}
\newcommand\sabBT{\mbox{$\sigma_{abB}(T)$}}
\newcommand\sabB{\mbox{$\sigma_{abB}$}}
\newcommand\sabTx{\mbox{$\sigma_{ab}(T,x)$}}
\newcommand\DsabTx{\mbox{$\Delta\sigma_{ab}(T,x)$}}
\newcommand\sabBTx{\mbox{$\sigma_{abB}(T,x)$}}

\newcommand\eg{{\it e.g.}}
\newcommand\ie{{\it i.e.}}
\newcommand\etal{{\it et al.}}

\newcommand\beq{\begin{\equation}}
\newcommand\eeq{\end{\equation}}

%---------------------------------------

\title{\mbox{}\vspace{2cm}\mbox{}\\
\Large\bf
The in-plane paraconductivity in \referee{\\}
\LSCOf\ thin film superconductors\referee{\\}
at high reduced-temperatures:\referee{\\}
Independence of the normal-state pseudogap\\ \mbox{}\\ \mbox{}\\ \mbox{}\\ }

\author{
\normalsize
Severiano R. Curr\'as,$^{1,2,*}$  Gonzalo Ferro,$^1$
M.~Teresa Gonz\'alez,$^1$\referee{\\ \normalsize}
Manuel V. Ramallo,$^1$ 
Mauricio  Ruibal,$^1$
Jos\'e Antonio Veira,$^1$\referee{\\ \normalsize}
Patrick Wagner,$^{1,2,\dagger}$ 
and F\'elix Vidal$^{1,\ddagger}$\\ \mbox{} \\ \normalsize
$^1$~LBTS,$^{\sharp}$ Departamento de F\'{\i}sica da Materia
Condensada,\referee{\\ \normalsize} Universidade de Santiago de Compostela E15782, Spain.\\ \normalsize
$^2$~Laboratorium voor Vaste-Stoffysica en Magnetisme, Katholieke\referee{\\ \normalsize} 
Universiteit Leuven, Celestijnenlaan 200 D, B3001 Heverlee, Belgium}

\date{}
\maketitle

\newpage

\begin{abstract}
The in-plane resistivity has been measured in
\LSCOf\ (\LSCO) superconducting thin films of underdoped ($x=0.10,0.12$),
optimally-doped ($x=0.15$) and overdoped ($x=0.20,0.25$) compositions. These films
were grown on (100)SrTiO$_3$ substrates, and have about 150~nm thickness. The in-plane  conductivity
induced by superconducting fluctuations above the superconducting
transition (the so-called in-plane paraconductivity, \Dsab) was
extracted from these data in the
reduced-temperature range
$10^{-2}\lsim\epsilon\equiv\ln(T/\Tc)\lsim1$. 
Such a \Dsabe\ was then analyzed in terms of 
 the mean-field--like Gaussian-Ginzburg-Landau (GGL) approach
extended to the high-$\epsilon$ region  by
means of the introduction of a total-energy cutoff, which takes
into account both the kinetic energy and the quantum localization
energy of each fluctuating mode. The obtained GGL coherence
length amplitude in the
$c$-direction, \xico, is constant  for
$0.10\leq x\leq0.15$ [$\xico\simeq0.9$~\AA], and  decreases with
increasing
$x$ in the overdoped range [$\xico\simeq0.5$~\AA\ for
$x=0.20$ and $\xico\sim0$~\AA\ for
$x=0.25$]. These results strongly suggest, therefore, that the
superconducting fluctuations in underdoped and overdoped \LSCO\ thin films may still be described, as in the
optimally-doped cuprates, in terms of the extended GGL
approach: The main effect of the doping  is just to change the
fluctuation dimensionality due to the change of the transversal
superconducting coherence length amplitude. In contrast, the
total-energy cutoff amplitude, \esuper, remains unchanged well within
the experimental uncertainties.    Our results
strongly suggest that at all temperatures above \Tc, including the
high reduced-temperature region, the doping mainly affects in \LSCO\ thin films the
normal-state properties and that its influence on the superconducting
fluctuations is relatively moderate: Even in the high-$\epsilon$
region, the in-plane paraconductivity is found to be independent of
the opening of a pseudogap in the normal state of the underdoped
films. We expect this last conclusion to be independent of the structural details of our films, \ie, applicable also to bulk samples.
\end{abstract}

\newpage

\section{Introduction}

It is now well established that some of the most central properties of
the high-tem\-perature cuprate superconductors (HTSC) strongly depend
on the hole doping.\cite{ReviewInicial} These
properties include, \eg, the normal-superconducting transition
temperature,
\Tc, and also the opening of a pseudogap in the normal region in
underdoped cuprates. Another  property much affected by doping is the
in-plane electrical conductivity
parallel to the CuO$_2$ planes, \sab, above the
normal-superconducting
transition.\cite{ReviewInicial} It is also known
that 
\sab\ in any HTSC is strongly affected around 
\Tc\ by the presence of evanescent Cooper pairs created by thermal
fluctuations (the so-called in-plane paraconductivity, \Dsab).
\cite{Tinkham,VidalRamallo} In fact, these fluctuation effects
may be appreciable even as far  above \Tc\ as
$T\simeq1.5\Tc$. So,
\sab\ may be decomposed as:
%%%%%%%%%%%%%%%%%%%%%%%%%%%%%%%%%%%%%%%%%%%%%%%%%%%%%%%%%%%%%%%%%%%%%%%%%%%%%%%%%%%%%%%%%%%%%%%%%%%%%%%%%%%%%%%%%%%%%%%%%%%%%%%%%%%%%%%%%%%%%%%%%%%%%%%%%%%%%%%%%%%%%%%%%%%%%%%
\begin{equation}
\sabTx=\DsabTx+\sabBTx,
\label{contribuciones}
\end{equation}
%%%%%%%%%%%%%%%%%%%%%%%%%%%%%%%%%%%%%%%%%%%%%%%%%%%%%%%%%%%%%%%%%%%%%%%%%%%%%%%%%%%%%%%%%%%%%%%%%%%%%%%%%%%%%%%%%%%%%%%%%%%%%%%%%%%%%%%%%%%%%%%%%%%%%%%%%%%%%%%%%%%%%%%%%%%%%%%
where $T$ is the temperature, $x$ the hole doping level, and
\sabBTx\ the so-called background or bare electrical in-plane
conductivity  (\ie, the electrical in-plane conductivity if the
fluctuations were absent).   It is then
natural to ask how much of the variation of
\sabTx\  observed when
the doping is changed is due to
\DsabTx,  and how much to \sabBTx. In fact, once \sabTx\ is measured,
the above question is equivalent to ask how \DsabTx\ is affected by
doping. This last question was first addressed by Suzuki and
Hikita\cite{Suzuki} and by Cooper and coworkers\cite{Cooper}, and
since then by different
authors\cite{Mori,Kimura,Juang,Asaka,Silva,Gueffaf,Meingast,Leridon,Varlamov}.
However, as of today some of the main conclusions (including, \eg,  the
dimensionality of the fluctuations or the influence of the normal-state pseudogap)
are still not well settled or they are even contradictory.  For instance, in the case of 
\YBCOf\ (\YBCO) some authors\cite{Cooper,Juang,Leridon} proposed that \Dsab\
becomes more three-dimensional (3D) when $x$ increases,
whereas other authors\cite{Mori,Gueffaf}  did not find any appreciable
dimensionality variation. Also, other authors have proposed, by  
analyzing either the fluctuation magnetization\cite{Varlamov} or the
thermal expansion\cite{Meingast},  that the superconducting
fluctuations in the underdoped \YBCO\ do not follow the Gaussian
mean-field--like theories used in
Refs.~\refcite{Cooper,Mori,Juang,Gueffaf,Leridon} but follow instead  different
forms of non-Gaussian  fluctuations.\cite{Meingast,Varlamov} 
Another example of these discrepancies is provided by the studies of 
\LSCOf\ (\LSCO): By analyzing their  measurements
of the 
 in-plane conductivity and magnetoconductivity, Suzuki and
Hikita\cite{Suzuki} proposed  a change in dimensionality
(2D to 3D) as doping increases from underdoped ($x<0.15$) to
overdoped ($x>0.15$). In contrast, 
 other authors\cite{Kimura} do not observe such a 
doping dependence in their  magnetorresistance measurements in the
same compounds.  Let
us stress here again that in addition to their interest for
understanding the superconducting fluctuations in HTSC, the dependence of the
paraconductivity on the doping   may also concern  other still open
problems, as the origin of the pseudogap which opens in the normal state
in the underdoped HTSC.\cite{ReviewInicial} For instance, various
theoretical models for such a pseudogap (see, \eg,
Refs.~\refcite{Randeria,Emery2D,EmeryStripes}) predict that the
fluctuation effects would strongly vary  with the doping level (even
changing order of magnitude), while in other models (see, \eg,
Refs.~\refcite{Pines,Bok,Anderson,Larkin}) the pseudogap  does not
result in a change in the superconducting fluctuations.

Among the
various possible reasons for the discrepancies commented above
between the different studies of \DsabTx, four of them seem to be
predominant: First, the different structural characteristics of the samples studied by each author (bulk or film samples, and in the latter case their thickness and substrate lattice parameters). In particular, as it is now well established,\cite{Chen,Loquet,Sato2,Bozovic,Kao} the substrate effects (including the associated strain effects) change some of the properties of these thin films, such as the absolute values of their  \Tc\ and in-plane normal resistivity. 
The {\it ``fine''} details of the paraconductivity, such as its dimensionality (which is directly related to the superconducting coherence length amplitude in the $c$-direction), may depend also on the structural nature of the samples. Let us stress already here, however, that it is also currently accepted that the most general aspects of the HTSC are similar for bulk  and thin film samples. This is the case, in particular, of the evolution of \Tc\ with doping or the appearance in the underdoped compositions of a pseudogap in the normal state.\cite{ReviewInicial,Suzuki,Cooper,Kimura,Loquet,Sato2,Kao} Therefore, the conclusions concerning the relationships between superconducting fluctuations and pseudogap effects or the existence or not of indirect paraconductivity effects (see below) may be expected to be general, \ie, independent of the structural characteristics of the superconductor.  A second source of ambiguity is the probable presence in some of the samples of
structural and stoichiometric inhomogeneities which, even when they
are relatively small, may appreciably affect the measured \sabTx,
mainly near \Tc.\cite{VidalInhomogeneidades}  Third, some
authors\cite{Suzuki,Gueffaf} analyze their \DsabTx\ data in
terms of the direct Aslamazov-Larkin (AL) contributions plus a
pair-breaking (Maki-Thompson, MT) term, while
others\cite{Mori,Juang,Asaka,Silva,Leridon} take into account only
the  AL contributions, and yet others\cite{Cooper,Kimura} do not
decide between both possibilities. Finally,  a fourth source of ambiguity is that the region of
reduced-temperatures,
$\epsilon\equiv\ln(T/\Tc)$, where the data are analyzed was relatively
small, typically $10^{-2}\lsim\epsilon\lsim0.1$, due to the fact
that the conventional GGL approach is not applicable very near 
or very far away from \Tc.\cite{Tinkham,VidalRamallo} To analyze the
region
$\epsilon\gsim0.1$, the usual
mean-field--like theories have to be extended to deal with
short-wavelength
fluctuations, which above \Tc\ become particularly
important when $\xi(T)$, the superconducting coherence
length, becomes of the order of \xio, its amplitude
extrapolated to $T$=0~K.\cite{Tinkham} Various attempts
to study the paraconductivity in the high-$\epsilon$ region as
a function of doping in HTSC have been recently
done by different authors.\cite{Asaka,Silva,Leridon} In particular,
Leridon \etal\cite{Leridon} analyzed the high-$\epsilon$
paraconductivity in \YBCO\ thin films with different dopings.
Unfortunately, these authors  based their analyses on a purely
heuristic expression for the high-$\epsilon$ paraconductivity, and
the physical meaning of the involved parameters   remains
unclear. The high-$\epsilon$ paraconductivity as a function of doping
in
\BSCOf\ was studied by Asaka
\etal\cite{Asaka} (using Ge-doped crystals) and by Silva
\etal\cite{Silva} (using oxygen-doped crystals but only in the
overdoped range). These authors  analyzed their data in terms of
the GGL approach with the conventional kinetic energy (also called
momentum) cutoff. Unfortunately, this conventional
cutoff\cite{Tinkham} extends the applicability of the GGL
paraconductivity only up to approximately
$\epsilon\simeq0.2$. Both groups find that the  
cutoff parameter is doping-dependent, what is attributed by Asaka
\etal\ to variations in the in-plane coherence length amplitude,
\xiabo, and by Silva \etal\ to possible deviations from
the BCS theory.

To further clarify the effects of doping  near \Tc\ on the
superconducting fluctuations in  HTSC, in this work we measure and
analyze the in-plane paraconductivity of different high-quality \LSCO\
thin films with thickness around 150~nm, grown on (100)SrTiO$_3$
 substrates and with Sr content corresponding to  $x$ = 0.10, 0.12,
0.15, 0.20 and 0.25. This kind of samples will allow us to compare even the {\it ``fine''} details of the paraconductivity measured in our work  with the measurements by  Suzuki and Hikita\cite{Suzuki}, done in \LSCO\ films grown on the same substrate as ours and with thickness $\sim 350$~nm.  We also take advantage of three important aspects with
respect to most of the previous works: First, the high controllability of the 
doping level in the \LSCO\ family has allowed us to grow 
$c$-axis oriented films with resistivities and
transition widths  among the best reported until now in
this
system.\cite{ReviewInicial,Suzuki,Loquet,Sato2,Kao}
Second, since the earlier results of
Refs.~\refcite{VeiraVidal,RamalloPRB} it is today well established
experimentally that the MT contributions to
\Dsab\ are negligible in optimally-doped HTSC. This
also agrees with the  calculations in
Refs.~\refcite{Yip,Maki} that indicate that for superconducting pairings with 
a $d$-wave component the strong pair-breaking effects of impurities
make negligible the MT terms. It was also demonstrated
experimentally  the absence in the in-plane paraconductivity
of optimally-doped HTSC of the so-called density-of-states (DOS)
contributions.\cite{RamalloPRB} So, in our analysis 
of the underdoped and overdoped \LSCO\ films we will assume  the absence
of appreciable indirect effects (MT and DOS). We are going to
see here that such an assumption is well confirmed by our
experimental results. A crucial advantage of our
present work is to use recent extensions of the
conventional mean-field--like calculations of
\Dsab\ to the short-wavelength regime
$\epsilon\gsim0.1$.\cite{CarballeiraDs,BriefReport}
These extensions are based on the introduction of a total-energy
cutoff in the  spectrum of the Gaussian-Ginzburg-Landau (GGL)
superconducting fluctuations, accounting for the Heisenberg
localization energy associated with the shrinkage, when the
reduced-temperature increases, of the superconducting
wave function.\cite{VidalEPL} These ``extended'' GGL expressions  have already
allowed us the analysis of the high-$\epsilon$ in-plane
paraconductivity of the optimally-doped
\YBCO,\cite{CarballeiraDs} and more recently of the optimally-doped
\BSCOf\ and
\Taliof.\cite{BriefReport} The high-$\epsilon$ paraconductivity in a
single underdoped \LSCO\ film, with
$x\simeq0.10$, was also briefly analyzed in Ref.~\refcite{BriefReport}.
Note, however, that in such work  the
superconducting fluctuations were assumed to be  essentially
2D. As we will see in the present work, this is not the
most likely scenario for our underdoped
\LSCO\ films in the
$\epsilon\lsim0.1$ region. In analyzing our data, we will clearly emphasize what results are expected to be independent of the structural nature of the samples and which ones concern the {\it ``fine''} behaviour of the paraconductivity and, therefore, are  applicable only to thin films.

In Sect.~II we describe the  samples' preparation
and  the  resistivity measurements. The
extraction from these data of the  in-plane paraconductivity, 
and their analysis in terms of the GGL model for superconducting
fluctuations with a total-energy cutoff, is presented in
Sect.~III. We summarize our
conclusions and discuss  their  main implications in Sect.~IV.

\section{Samples' preparation \protect\referee{\protect\\} and
electrical
 resistivity measurements}

The samples studied in this work are $c$-axis--oriented \LSCO\ thin
films grown on (100)SrTiO$_3$  substrates from ceramic
single-targets with different Sr contents.  All the films
have similar thickness, of around 150~nm (see Table~I).
During deposition, by high-pressure DC sputtering with an on-axis
cathode-substrate configuration, the substrate was held at
temperatures between 840-860~\mbox{$^{\rm o}$C} in a flow of pure
oxygen at 1.3 Torr.  After deposition, the films were maintained
during 30 minutes at  the same temperature but at oxygen pressure of
7-10 Torr. Then, they were cooled down to room temperature to ensure
full oxigenation.  The crystal structure
of the \LSCO\ films was studied by X-ray diffraction in
a Bragg-Brentano geometry. The diffraction spectra exhibit only
(00$l$) peaks, indicating then an oriented growth with
the
$c$-axis perpendicular to the substrate. The full-width at
half-maximum of the rocking curve corresponding to the (006)
reflection is around 0.3$^{\rm o}$ for all the studied
samples.  Films were then patterned as narrow strips,  with typical
widths from 5 to 10 $\mu$m and typical  lengths of 100 $\mu$m, by
photolithography and wet chemical etching. Then, Au contact pads
were deposited onto the current and voltage terminals, and annealed
in oxygen at 1 atm and 600~\mbox{$^{\rm o}$C} during 15 minutes to
facilitate gold diffusion into the \LSCO. The final resistance
achieved was less than 0.1~$\Omega$ per contact. 
The in-plane
resistivity of the films,
\rhoabT, was then measured by using a standard four-contact DC
current arrangement in the temperature range from 4.2~K up to 300~K.
The applied current  was $\sim$5~\mbox{$\mu$A}, which implies
that the current density through the sample is around
100~\mbox{Acm$^{-2}$} (much lower than the
critical current density in zero applied magnetic field). The
resolutions in the
\rhoabT\ measurement are about  1~\mbox{$\mu\Omega$cm} for
resistivity, and about  10~mK for temperature.

In Figs.~\figuno\ to \figtres\ it is shown the temperature-dependence of
\rhoabT\  for our
\LSCO\ thin films, with Sr contents $x$=0.10, 0.12, 0.15, 0.20 and
0.25.  These $x$ values  are the ones given by the nominal
deposition rates. As it can be seen in these figures, upon increasing
the Sr content the in-plane resistivity decreases gradually and its
$T$-dependence  well above \Tc\ changes systematically from a
concave shape (for $x\leq0.15$) to a more linear one (for $x>0.15$).
Such a  doping-dependence of the resistivity  is in good agreement with
previous results in similar \LSCO\
films.\cite{ReviewInicial,Suzuki,Loquet,Sato2,Kao,BriefReport} For all  the films
studied here, the values of \rhoab\ at 250~K (see Table~I) are among the lowest
reported until now in the literature for films with the same doping and substrate
and similar thickness.\cite{ReviewInicial,Suzuki,Loquet,Sato2,Kao,BriefReport} 
Another indication of the quality of our samples is the width of the superconducting
transition in the
\rhoabT\ curves, which are also among the smallest in the literature.
In Fig.~\figtres\ it is shown the detail of
\rhoabT\ around the superconducting transition  and the corresponding
d\rhoab/d$T$. In this figure it is also shown \Tci, the temperature
where d\rhoab/d$T$ reaches its maximum, and \Tcim\ and \Tcip, 
corresponding to the half-maximum of such a derivative peak above and,
respectively, below \Tci. The value of
\Tci, \Tcim\ and \Tcip\ for all  the samples studied in this work is
given in  Table~I. Also, in Fig.~\figTci\ we show the \Tci\
values of our films as a function of Sr content.  As it can be easily
seen in such figure,
\Tci\ rises with increasing the Sr content
 up to a optimal doping level of about $x$ = 0.15, and then
decreases. These \Tci\ values,
and also their doping dependence,  are in good agreement with the
ones at present well established for
\Tc\ in \LSCO\ films with similar substrate and
thickness (note that the \Tci\ values of good-quality bulk \LSCO\ are higher than in thin films with similar Sr content, although both types of samples share the same \Tci\ dependence on doping).\cite{ReviewInicial,Suzuki,Loquet,Sato2,Kao,BriefReport}
In the remaining of this work we will use, unless
specified otherwise, \Tci\
as \Tc. The appropriateness of such a choice in the
analysis of the in-plane paraconductivity  will be  discussed
in further detail in subsection~{\ref{secuncertainties}}.

\section{The in-plane paraconductivity:
Comparison with the extended
GGL approach}

\subsection{Extraction of the in-plane paraconductivity}
As
usually,\cite{Tinkham,VidalRamallo}
\Dsab\ is obtained from the measured \roabT\ curves by just using
Eq.\eq{contribuciones} and estimating the normal-state background
$\roabB$ (\ie, $\sigma_{abB}^{-1}$)  by extrapolating through the
transition the
\roabT\ data measured well above the region where \Dsabe\ is to be
analyzed. As we want to analyze \Dsabe\ also in the high-$\epsilon$
($\epsilon>0.1$) region, we will use a procedure similar to the one
already used in Refs.~\refcite{CarballeiraDs,BriefReport}. This is a
two-step process: First, we fit the \roabT\ data in the
$T$-region $1.6\,\Tc\lsim T\lsim2.7\,\Tc$ (corresponding to
$0.5\lsim\epsilon\lsim1$). To perform such a fit, we
used the simplest functionality producing good agreement with the data
at all dopings, which is a linear  plus a Curie-like term, $a/T+b+cT$.
Indeed, such a  background is appropriate to analyze
\Dsabe\ only well below $\epsilon\simeq0.5$. And, in fact, similar
types of extrapolations have been successfully used to obtain \Dsabe\
up to
$\epsilon\simeq0.1$ in various optimally-doped
HTSC (see, \eg, Refs.~\refcite{VidalRamallo,VeiraVidal,RamalloPRB,Pavuna}).
Now then, by construction this background makes
\Dsab\ to become zero at
$\epsilon=0.5$, and thus cannot lead to a trustworthy \Dsabe\
analysis for the high-$\epsilon$ ($\epsilon>0.1$) region. So, the
second step to determine a background valid for our purposes will
be to fit the
\roabT\ data at considerably higher reduced-temperatures, in the
$T$-range
$4.5\,\Tc\lsim T\lsim7\,\Tc$ (which corresponds to
$1.5\lsim\epsilon\lsim2$). However, because the
extrapolation uncertainties strongly increase with the $T$-distance,
we require to this last fit to reproduce   in the
moderate-$\epsilon$ range 
$10^{-2}\lsim\epsilon\lsim0.1$ the \Dsabe\ results obtained with
our first background estimate, up to a $\pm20\%$ maximum uncertainty.
We used for these fits a Curie-like plus a 2nd.-degree polynomial 
term,
$a/T+b+cT+dT^2$, which is  the simplest functional producing in
the enlarged $T$-region good fits  for all $x$. A detailed account of
the uncertainty associated with our background estimation, mainly
focused on its impact on our analyses of \Dsabe, will be presented
later in subsection~\ref{secuncertainties}. However, let us emphasize
already here that such backgrounds do not
appreciably depend on variations of the lower limit of the
background fitting region, $\TLB$, from its default value
$4.5\Tc$, provided that \TLB\  is kept above $\sim3.5\Tc$ (\ie,
above $\epsilon\simeq1.2$). In Figs.~\figuno\ to \figtres\ we  plot the
normal-state backgrounds obtained by using the above procedure in
all  the films studied in this work. Note  that the
upwards concavity of
\roabBT\ diminishes for
$x\geq0.15$. Also shown in Figs.~\figuno\ and \figdos\ is \Tsuper, the
temperature at which \Dsabe\ is observed to become negligible. We
emphasize
 that \Tsuper\ is  well below \TLB, and also that 
there is few variation of \Tsuper\ when \TLB\ is moved, always if
$\TLB\gsim3.5\Tc$ (see also subsection~\ref{secuncertainties}). This
indicates that the existence of such a
\Tsuper\ is not an artifact of the background subtraction procedure.

In Fig.~\figcuatro\ we show the \Dsab-versus-$\epsilon$ curves
obtained for each of the films measured in this work. This figure
already illustrates a  result central in   our present paper: The
experimental
$\Dsab(\epsilon)$ curves agree with each other well within the
experimental errors for the doping levels
$0.10\leq x\leq0.15$ (corresponding to the underdoped and
optimally-doped range of compositions). Note that this conclusion
does not rely on any comparison with any theory. This striking result
already suggests   the non-validity, at least  for
\LSCO\ films, of the proposals  by various authors (see, \eg,
Refs.~\refcite{ReviewInicial,Meingast,Varlamov,Randeria,Emery2D,EmeryStripes})
that the superconducting fluctuations are substantially different in
underdoped and optimally-doped HTSC. As also shown
in Fig.~\figcuatro, in the overdoped range of
compositions (\ie, for  $x=0.20$ and 0.25) the in-plane
paraconductivity of the \LSCO\ films increases with the value of
$x$, mainly in the $\epsilon$-region $\epsilon\lsim0.1$. At higher
reduced-temperatures, all  the $\Dsab(\epsilon)$ curves (for all
$x$) collapse towards negligible paraconductivity at a
reduced-temperature $\epsilon$ of around 0.8.

\subsection{Theoretical background: 
 Extension of the GGL paraconductivity  to the high-$\epsilon$ region
\label{teorbackg}}

To analyze the experimental data summarized in the above
subsection, we will use the paraconductivity expressions obtained
on the grounds of the mean-field GGL approach regularized through the
so-called ``total-energy'' cutoff, which takes into account the limits imposed by the uncertainty principle to the shrinkage of the superconducting wave function when the temperature increases well above \Tc.\cite{VidalEPL} Our results in optimally-doped HTSC suggest that such a regularization extends the applicability of the mean-field--like GGL approach from the $\epsilon_{\rm LG}\lsim\epsilon\lsim0.1$ region to the high-$\epsilon$
region.\cite{CarballeiraDs,BriefReport,VidalEPL} Here $\epsilon_{\rm LG}$ is the so-called Levanyuk-Ginzburg
reduced-temperature, below which the fluctuations enter in the
full-critical, non-Gaussian region where non-mean-field approaches (like the 3DXY model) must be applied.\cite{eLG} Such an $\epsilon_{\rm LG}$ was estimated to be of the order of $10^{-2}$ in HTSC.\cite{VidalRamallo,eLG} Note also  that so close to \Tc\ the effects of sample inhomogeneities may considerably affect the paraconductivity data.\cite{VidalInhomogeneidades} Therefore, in the present paper we will will restrict our analyses  to the  mean-field--like $\epsilon\gsim10^{-2}$ region. Although the GGL in-plane
paraconductivity under the total-energy cutoff was calculated for the first time in
Ref.~\refcite{CarballeiraDs}, it will be useful to summarize in this subsection some
of the results of such calculations for the case of single-layered superconductors,
and also remember the physical meaning of the total-energy cutoff, which was
analyzed in terms of the uncertainty principle applied to the superconducting wave
function in Ref.~\refcite{VidalEPL}. For layered superconductors with a single
interlayer separation, $s$ (the case of the \LSCO\ family, where $s=6.6$~\AA), the
total-energy cutoff may be written as:
%%%%%%%%%%%%%%%%%%%%%%%%%%%%%%%%%%%%%%%%%%%%%%%%%%%%%%%%%%%%%%%%%%%%%%%%%%%%%%%%%%
\begin{equation}
k_{xy}^2+
{{\BLD(1-\cos(k_z s))}\over{2\xi_{ab}^{2}(0)}}+
\xi_{ab}^{-2}(\epsilon)
\,\leq\,
\xi_{ab0}^{-2}.\label{cutoff}
\end{equation}
%%%%%%%%%%%%%%%%%%%%%%%%%%%%%%%%%%%%%%%%%%%%%%%%%%%%%%%%%%%%%%%%%%%%%%%%%%%%%%%%%%
In this expression,  $\mbox{\bf k}_{xy}$ and $\mbox{\bf k}_z$ are
the in-plane and
$c$-direction wavevectors of the fluctuating modes,
$\xiabe=\xiabo\epsilon^{-1/2}$ is the in-plane GL  coherence
length, $\BLD\equiv(2\xico/s)^2$ is the so-called Lawrence-Doniach
(LD) dimensional crossover parameter, \xiabo\ and  \xico\ are the in-plane and, respectively, out-of-plane 
GL superconducting coherence  length  amplitudes,  and $\xi_{ab0}$ is
the in-plane Pippard coherence length.  Note   that  $\mbox{\bf
k}_z$  is limited by the layered structure as 
$-\pi/s\leq k_z\leq\pi/s$.

The left-hand side in Eq.\eq{cutoff}  is the total energy of a
fluctuation mode (in units of $\hbar^2/2m_{ab}^*$, where
$\hbar$ is the reduced Plank constant and  $m_{ab}^*$ is the in-plane
effective mass of the superconducting pairs). As explained in
Ref.~\refcite{VidalEPL}, this ``total-energy'' of each fluctuation
mode may be seen as the sum of the Heisenberg localization energy
associated   with the  shrinkage of
the superconducting wave function when the temperature increases
above \Tc\ [the 
$\xi_{ab}^{-2}(\epsilon)$ term] and the conventional kinetic energy.
In the case of single-layered superconductors, the kinetic energy
appears as the sum of two contributions: The in-plane kinetic
energy (the $k_{xy}^2$ term) and the
$c$-direction kinetic energy savings [the $\BLD(1-\cos(k_z
s))/2\xi_{ab}^{2}(0)$ term]. The right-hand side in
Eq.\eq{cutoff} may be seen as the localization energy 
associated to the
maximum shrinkage, at $T=0$~K, of the superconducting
wave function. This term is, therefore, proportional to the inverse
square of the in-plane Pippard coherence length amplitude, 
$\xi_{ab0}$ (see Ref.~\refcite{VidalEPL}).

To briefly analyze the differences between the total-energy
cutoff and the conventional momentum or kinetic-energy cutoff,
the simplest case is the 2D layered limit, where $\BLD=0$. In
that case Eq.\eq{cutoff} simplifies to:
%%%%%%%%%%%%%%%%%%%%%%%%%%%%%%%%%%%%%%%%%%%%%%%%%%%%%%%%%%%%%%%%%%%%%%%%%%%%%%%%%%
\begin{equation}
k_{xy}^2+
\xi_{ab}^{-2}(\epsilon)
\,\leq\,
\xi_{ab0}^{-2}.\label{cutoff2D}
\end{equation}
%%%%%%%%%%%%%%%%%%%%%%%%%%%%%%%%%%%%%%%%%%%%%%%%%%%%%%%%%%%%%%%%%%%%%%%%%%%%%%%%%%
Near
\Tc, when
$\xiab(\epsilon)\gg\xi_{ab0}$, the localization energy contribution
to each fluctuation mode may be neglected, and Eq.\eq{cutoff2D}
reduces then to the conventional momentum or kinetic-energy
cutoff in 2D layered superconductors:\cite{Tinkham}
%%%%%%%%%%%%%%%%%%%%%%%%%%%%%%%%%%%%%%%%%%%%%%%%%%%%%%%%%%%%%%%%%%%%%%%%%%%%%%%%%%
\begin{equation}
k_{xy}^2
\,\leq\,
c\,\xi_{ab}^{-2}(0).\label{kecutoff2D}
\end{equation}
%%%%%%%%%%%%%%%%%%%%%%%%%%%%%%%%%%%%%%%%%%%%%%%%%%%%%%%%%%%%%%%%%%%%%%%%%%%%%%%%%%
where instead of the Pippard $\xi_{ab0}$ we have used
$c^{-1/2}\xiabo$, where $c$ is a cutoff amplitude,
temperature-independent, close to 1.  These expressions  still
simplify in the case of isotropic 3D superconductors: The
total-energy cutoff reduces to $k^2+
\xi^{-2}(\epsilon)
\,\leq\,
\xi_{0}^{-2}$, whereas one refinds $k^2 
\,\leq\,
c\,\xi^{-2}(0)$ for the momentum or kinetic-energy cutoff, which is
the familiar condition earlier proposed for low-\Tc\
superconductors.\cite{Tinkham}
By using again $\xi_{0}=c^{-1/2}\xi(0)$ and assuming the
applicability at all reduced-temperatures of the mean-field
$\epsilon$-dependence of the superconducting coherence length,
$\xi(\epsilon)=\xi(0)\epsilon^{-1/2}$, one may see that the conventional
kinetic-energy
and the  total-energy  cutoffs are
related, in the 2D and 3D limits,  through the substitution of $c$ by
$c-\epsilon$. So, as stressed before both cutoffs coincide  near
\Tc, when
$\epsilon\ll c$. The conventional momentum or kinetic-energy cutoff appears
then as a particular case, the limit when $\xi(\epsilon)\gg\xi_0$, of the
total-energy cutoff. However, in spite of the simple relationship between both
cutoff approaches, first proposed in Ref.~\refcite{MosqueiraPRL}, their deep
conceptual differences also lead to striking
differences in the high-$\epsilon$ behaviour of any observable
associated with the superconducting fluctuations above
\Tc, included the paraconductivity. These differences have been
analyzed in Refs.~\refcite{CarballeiraDs,BriefReport,VidalEPL}, but it
could be useful to stress here some of them. Note first that the
maximum in-plane kinetic energy of the fluctuating Cooper pairs,
$E_{ab,\rm kinetic}^{\,\rm max}$, is temperature-independent in the
case of the conventional  kinetic-energy or momentum cutoff. For
instance, for 2D layered superconductors,
%%%%%%%%%%%%%%%%%%%%%%%%%%%%%%%%%%%%%%%%%%%%%%%%%%%%%%%%%%%%%%%%%%%%%%%%%%%%%%
\begin{equation}
E_{ab, \rm
kinetic}^{\,\rm max} ({\rm momentum \ cutoff})=
{{(\hbar k_{xy}^{\rm max} )^2}\over{2m_{ab}^*}}=
{{\hbar^2}\over{2m_{ab}^*\xi_{ab}^2(0)}}\, c\, .\label{kemaxmom2D}
\end{equation}
%%%%%%%%%%%%%%%%%%%%%%%%%%%%%%%%%%%%%%%%%%%%%%%%%%%%%%%%%%%%%%%%%%%%%%%%%%%%%%
In contrast, under a total-energy cutoff the maximum in-plane
kinetic energy of the Cooper pairs is temperature-dependent. In this
example (2D layered superconductors), the corresponding
$E_{ab, \rm kinetic}^{\,\rm max}$ may be directly obtained from
Eq.\eq{kemaxmom2D} by using $c-\epsilon$ instead of $c$ [or from
Eq.\eq{cutoff2D} and using again $\xi_{ab0}=c^{-1/2}\,\xi_{ab}(0)$ and
$\xi_{ab}(\epsilon)=\xi_{ab}(0)\epsilon^{-1/2}$] as:
%%%%%%%%%%%%%%%%%%%%%%%%%%%%%%%%%%%%%%%%%%%%%%%%%%%%%%%%%%%%%%%%%%%%%%%%%%%%%%
\begin{equation}
E_{ab, \rm
kinetic}^{\,\rm max} (\mbox{\rm total-energy  cutoff})=
{{\hbar^2}\over{2m_{ab}^*\xi_{ab}^2(0)}}\,(c-\epsilon),
\;\;\;\;\;\;\;\;{\rm with }\;\epsilon\leq c,
\label{kemaxte2D}
\end{equation}
%%%%%%%%%%%%%%%%%%%%%%%%%%%%%%%%%%%%%%%%%%%%%%%%%%%%%%%%%%%%%%%%%%%%%%%%%%%%%%
which is temperature-dependent and that becomes zero for
$\epsilon\geq c$. In other words, in contrast with the conventional
momentum or kinetic-energy cutoff which only eliminates,
independently of the temperature, the fluctuating modes with
in-plane kinetic-energy above $c\hbar^2/2m_{ab}^*\xi_{ab}^2(0)$, the
total-energy cutoff eliminates {\it all} the fluctuation modes
at reduced-temperatures equal to  $c$ or above. By
imposing a zero kinetic-energy in  Eqs.~\eq{cutoff} or \eq{cutoff2D},
this reduced-temperature, denoted
\esuper, is given by:
%%%%%%%%%%%%%%%%%%%%%%%%%%%%%%%%%%%%%%%%%%%%%%%%%%%%%%%%%%%%%%%%%%%%%%%%%%%%%%
\begin{equation}
\xi_{ab}(\esuper)=\xi_{ab0},
\label{obtenerec}
\end{equation}
%%%%%%%%%%%%%%%%%%%%%%%%%%%%%%%%%%%%%%%%%%%%%%%%%%%%%%%%%%%%%%%%%%%%%%%%%%%%%%
\ie,  
$\epsilonsuper=(\xiabo/\xiabsubo)^2$. As first argued in
Ref.~\refcite{VidalEPL}, Eq.\eq{kemaxte2D} (which leads directly to the
existence of a well-defined reduced-temperature above which all
coherent Cooper pairs vanish) may be seen as just a 
consequence of the limitations imposed by the uncertainty principle to the
shrinkage of the superconducting wave function, which also above
\Tc\ imposes the condition $\xi_{ab}(\epsilon)\geq\xi_{ab0}$: In
other words, the collective behaviour of the Cooper pairs will be
dominated at high reduced-temperatures by the Heisenberg localization
energy.\cite{VidalEPL} If, in addition, we assume the applicability
of the BCS relationship in the clean limit, 
$\xiabo=0.74\,\xiabsubo$, then
$\epsilon^C_{\rm BCSclean}=c_{\rm
BCSclean}\simeq0.6$.\cite{BriefReport,VidalEPL,notadirty} Indeed this
value of
$c$ will also apply to the conventional momentum cutoff approach
that appears as the limit when $\epsilon\ll 1$ of the total-energy
cutoff.

As we are particularly interested in analyzing the dimensionality of
the superconducting fluctuations in \LSCO\ as a function of the
doping, we will first summarize here the general expressions of the
paraconductivity as a function of the LD dimensional crossover
parameter, \BLD. Then we will also present the limiting cases
$\BLD\ll\epsilon$ (2D limit) and $\BLD\gg\epsilon$ (3D limit). The
paraconductivity  for single-layered superconductors resulting
from the GGL approach extended to high reduced-temperatures by the total-energy cutoff
 has been calculated in Ref.~\refcite{CarballeiraDs} to be:
%%%%%%%%%%%%%%%%%%%%%%%%%%%%%%%%%%%%%%%%%%%%%%%%%%%%%%%%%%%%%%%%%%%%%%%%%%%%%%%%%%
\begin{equation}
\Delta\sigma_{ab}(\epsilon)_{\rm E}={{\mbox{e}^2}\over{16\hbar
s}}
\left[
{{1}\over{\epsilon}}
\left(
1+{{\BLD}\over{\epsilon}}
\right)^{-1/2}
-
{{1}\over{\epsilonsuper}}\left(2-
{{\epsilon+\BLD/2}\over{\epsilonsuper}}
\right)\right],\label{teor}
\end{equation}
%%%%%%%%%%%%%%%%%%%%%%%%%%%%%%%%%%%%%%%%%%%%%%%%%%%%%%%%%%%%%%%%%%%%%%%%%%%%%%%%%%
where  e
is the electron's charge.  The paraconductivity
under a total-energy cutoff in the 2D limit may be then obtained by
just applying in Eq.~\eq{teor} the condition
$\BLD\ll\epsilon$:\cite{CarballeiraDs}
%%%%%%%%%%%%%%%%%%%%%%%%%%%%%%%%%%%%%%%%%%%%%%%%%%%%%%%%%%%%%%%%%%%%%%%%%%%%%%%%%%
\begin{equation}
\Delta\sigma_{ab}^{\rm 2D}(\epsilon)_{\rm
E}={{\mbox{e}^2}\over{16\hbar s}}
\left[
{{1}\over{\epsilon}}
-
{{1}\over{\epsilonsuper}}\left(2-
{{\epsilon}\over{\epsilonsuper}}
\right)\right].\label{teor2D}
\end{equation}
%%%%%%%%%%%%%%%%%%%%%%%%%%%%%%%%%%%%%%%%%%%%%%%%%%%%%%%%%%%%%%%%%%%%%%%%%%%%%%%%%%
Concerning the paraconductivity under a total-energy cutoff in the 3D
limit, it cannot be directly obtained from Eq.\eq{teor} because this
equation assumes the $c$-direction layered-structure cutoff,
$-\pi/s\leq k_z\leq\pi/s$, which in the limit $\BLD\gg\epsilon$ is no
longer a stronger limitation for $k_z$ than the total-energy cutoff
[Eq.\eq{cutoff}].\cite{CarballeiraDs} The calculations using  the
total-energy cutoff for the three directions of space have been also
done in Ref.~\refcite{CarballeiraDs}, the  resulting expression being:
%%%%%%%%%%%%%%%%%%%%%%%%%%%%%%%%%%%%%%%%%%%%%%%%%%%%%%%%%%%%%%%%%%%%%%%%%%%%%%%%%%
\begin{equation}
\Delta\sigma_{ab}^{\rm 3D}(\epsilon)_{\rm
E}=
{{\mbox{e}^2}\over{48\pi\hbar\xi(0)}}
\left\{
3\left[
{{{\rm arctan}(\sqrt{(\esuper-\epsilon)/\epsilon})}\over{\sqrt{\epsilon}}}
-
{{\epsilon\sqrt{\esuper-\epsilon}}\over{(\esuper)^2}}
\right]
-5
{{(\esuper-\epsilon)^{3/2}}\over{(\esuper)^2}}
\right\}.\label{teor3D}
\end{equation}
%%%%%%%%%%%%%%%%%%%%%%%%%%%%%%%%%%%%%%%%%%%%%%%%%%%%%%%%%%%%%%%%%%%%%%%%%%%%%%%%%%

Let us also summarize the results for  the in-plane
paraconductivity under the conventional momentum
cutoff.  As first explicitly derived by Asaka \etal\cite{Asaka} (see also
Ref.~\refcite{CarballeiraDs} for a more detailed calculation), the
corresponding expression for a single-layered
superconductor is:  
%%%%%%%%%%%%%%%%%%%%%%%%%%%%%%%%%%%%%%%%%%%%%%%%%%%%%%%%%%%%%%%%%%%%%%%%%%%%%%%%%%
\begin{equation}
\Delta\sigma_{ab}(\epsilon)_{\rm M}=
{{\mbox{e}^2}\over{16\hbar s}}
\left\{
{{1}\over{\epsilon}}
\left(
1+{{\BLD}\over{\epsilon}}
\right)^{-1/2}\hskip-2em
-
{{c(c+\epsilon+\BLD/2)}
\over
{[(c+\epsilon+\BLD)(c+\epsilon)]^{3/2}}}
-
{{1}\over{\epsilon+c}}
\left(
1+{{\BLD}\over{\epsilon+c}}
\right)^{-1/2}
\right\}.\label{teorM}
\end{equation}
%%%%%%%%%%%%%%%%%%%%%%%%%%%%%%%%%%%%%%%%%%%%%%%%%%%%%%%%%%%%%%%%%%%%%%%%%%%%%%%%%%
The corresponding results in the 2D and 3D limits are,
respectively:
%%%%%%%%%%%%%%%%%%%%%%%%%%%%%%%%%%%%%%%%%%%%%%%%%%%%%%%%%%%%%%%%%%%%%%%%%%%%%%%%%%
\begin{equation}
\Delta\sigma_{ab}^{2D}(\epsilon)_{\rm M}=
{{\mbox{e}^2}\over{16\hbar s}}
\left[
{{1}\over{\epsilon}}
-
{{c}\over{(c+\epsilon)^2}}
-
{{1}\over{\epsilon+c}}
\right], \label{teorM2D}
\end{equation}
%%%%%%%%%%%%%%%%%%%%%%%%%%%%%%%%%%%%%%%%%%%%%%%%%%%%%%%%%%%%%%%%%%%%%%%%%%%%%%%%%%
and
%%%%%%%%%%%%%%%%%%%%%%%%%%%%%%%%%%%%%%%%%%%%%%%%%%%%%%%%%%%%%%%%%%%%%%%%%%%%%%%%%%
\begin{equation}
\Delta\sigma_{ab}^{\rm 3D}(\epsilon)_{\rm
M}=
{{\mbox{e}^2}\over{48\pi\hbar\xi(0)}}
\left\{
3\left[
{{{\rm arctan}(\sqrt{c/\epsilon})}\over{\sqrt{\epsilon}}}
-
{{\epsilon\sqrt{c}}\over{(\epsilon+c)^2}}
\right]
-5
{{c^{3/2}}\over{(\epsilon+c)^2}}
\right\}.\label{teorM3D}
\end{equation}
%%%%%%%%%%%%%%%%%%%%%%%%%%%%%%%%%%%%%%%%%%%%%%%%%%%%%%%%%%%%%%%%%%%%%%%%%%%%%%%%%%
Equations \eq{teorM2D} and \eq{teorM3D} correspond to the
expressions first obtained for these 2D and 3D limits by Gauzzi and
Pavuna in Ref.~\refcite{Pavuna}. It is also useful to note here the differences between the asymptotic behaviour of \Dsabe\ under the two different cutoff conditions. For instance, as it is well known,\cite{CarballeiraDs,Pavuna} in the 2D limit the conventional momentum cutoff predicts that $\Delta\sigma_{ab}^{\rm 2D}(\epsilon)_{\rm M}$ smoothly decays as $\epsilon^{-3}$ when $\epsilon\gg c$. In contrast, as stressed before,  the paraconductivity under a total-energy cutoff presents a singularity when $\epsilon=\esuper=c$. Such a behaviour is not describable through a critical exponent in $\epsilon$. However, when $\epsilon$ approaches $c$ from below such a singular behaviour may be described in terms of a power law in $|\tilde\epsilon|$, where $\tilde\epsilon\equiv\epsilon-c$. In the 2D limit, this asymptotic behaviour is: 
%%%%%%%%%%%%%%%%%%%%%%%%%%%%%%%%%%%%%%%%%%%%%%%%%%%%%%%%%%%%%%%%%%%%%%%%%%%%%%%%%%
\begin{equation}
\Delta\sigma_{ab}^{\rm 2D}(\tilde\epsilon)_{\rm E}={{{\rm e}^2 }\over{16\hbar s c^3}} |\tilde\epsilon|^2, \;\;\;\;\; {\rm for }\;\;\tilde\epsilon\rightarrow0^-.
\end{equation}
%%%%%%%%%%%%%%%%%%%%%%%%%%%%%%%%%%%%%%%%%%%%%%%%%%%%%%%%%%%%%%%%%%%%%%%%%%%%%%%%%%

The in-plane paraconductivity without any cutoff may be directly
obtained from the above \Dsabe\ expressions by
just imposing 
$\epsilon\ll\esuper$ in Eqs.\eq{teor} to \eq{teor3D} [or $\epsilon\ll
c$ in Eqs.\eq{teorM} to \eq{teorM3D}]. This leads to the well-known 
Lawrence-Doniach expression for a single-layered superconductor:
%%%%%%%%%%%%%%%%%%%%%%%%%%%%%%%%%%%%%%%%%%%%%%%%%%%%%%%%%%%%%%%%%%%%%%%%%%%%%%%%%%
\begin{equation}
\Delta\sigma_{ab}(\epsilon)_{\rm
no\,cutoff}={{\mbox{e}^2}\over{16\hbar s}} {{1}\over{\epsilon}}
\left(
1+{{\BLD}\over{\epsilon}}
\right)^{-1/2},\label{teorLD}
\end{equation}
%%%%%%%%%%%%%%%%%%%%%%%%%%%%%%%%%%%%%%%%%%%%%%%%%%%%%%%%%%%%%%%%%%%%%%%%%%%%%%%%%%
which recovers in the limits $\BLD\ll\epsilon$ and $\BLD\gg\epsilon$
the also well-known results by Aslamazov and Larkin for the
paraconductivity without any cutoff in 2D and 3D superconductors:
%%%%%%%%%%%%%%%%%%%%%%%%%%%%%%%%%%%%%%%%%%%%%%%%%%%%%%%%%%%%%%%%%%%%%%%%%%%%%%%%%%
\begin{equation}
\Delta\sigma_{ab}^{\rm 2D}(\epsilon)_{\rm no\,cutoff}=
{{\mbox{e}^2}\over{16\hbar
s\epsilon}},\label{teorLD2D}
\end{equation}
%%%%%%%%%%%%%%%%%%%%%%%%%%%%%%%%%%%%%%%%%%%%%%%%%%%%%%%%%%%%%%%%%%%%%%%%%%%%%%%%%%
and
%%%%%%%%%%%%%%%%%%%%%%%%%%%%%%%%%%%%%%%%%%%%%%%%%%%%%%%%%%%%%%%%%%%%%%%%%%%%%%%%%%
\begin{equation}
\Delta\sigma_{ab}^{\rm 3D}(\epsilon)_{\rm no\,cutoff}=
{{\mbox{e}^2}\over{32\hbar
\xi_c(0)}} \epsilon^{-1/2}.\label{teorLD3D}
\end{equation}
%%%%%%%%%%%%%%%%%%%%%%%%%%%%%%%%%%%%%%%%%%%%%%%%%%%%%%%%%%%%%%%%%%%%%%%%%%%%%%%%%%

Finally, let us stress here that all  the above expressions for
the in-plane paraconductivity [Eqs.\eq{teor} to \eq{teorLD3D}]
implicitly assume a relaxation
time for the superconducting fluctuations equal to the one given by
the standard BCS mean-field approach,
$\tau_0=(\pi\hbar/8\kB\Tc)\epsilon^{-1}$.\cite{tau} They also consider that 
 all indirect contributions (as the Maki-Thompson and the
density-of-states ones) to \Dsabe\ are negligible, as it is today
well established for the HTSC.\cite{VeiraVidal,RamalloPRB}

\subsection{Comparison between the experimental data and
the  GGL  paraconductivity}

We present
in  Fig.~\figcuatro\ a comparison between  the
theoretical in-plane paraconductivity under a total-energy cutoff
  and the experimental data obtained in the different
\LSCO\ films measured in this work. The solid lines in these figures are the best
fits of Eq.\eq{teor} to the experimental data in all the measured
$\epsilon$-region. In these fits, we have used
\Tci\ as \Tc, the only free parameters being 
\xico\ and
\epsilonsuper. The so-obtained values for \xico\ and \esuper\ are
summarized in Table~I. The error bars and the $x$-dependence of
these values will be discussed later in detail. As it is
easily observable in this Fig.~\figcuatro, the agreement between Eq.\eq{teor} and
the experimental data is excellent in all the region
$10^{-2}\lsim\epsilon\lsim 1$, for all the samples studied in
this work. This comparison confirms, in particular, 
the existence of a well-defined reduced-temperature, \esuper, of about
$\esuper\sim0.8$, above which  the in-plane paraconductivity vanishes. 
A different view of these fits  is provided by
Figs.~\figdos\ and \figtres. In these figures, the continuous lines
are the theoretical \roabT\ curves which
result from the above analyses [\ie, from adding, through
Eq.\eq{contribuciones}, the normal-state background and the 
theoretical \Dsabe\ resulting from the above fits]. As visible in
these figures, the agreement   with the  \roabT\  measurements
above the transition is again excellent.

It may be useful to check if the GGL theory with a total-energy
cutoff may also explain the in-plane resistivity 
measurements done by Suzuki and Hikita\cite{Suzuki}
in their own
\LSCO\ films, with various
$x$ values, grown on the same substrate as ours, and with 
thickness $\sim350$~nm. For that, we have scanned the corresponding
published
\roabT\ plots\cite{Suzuki} and
applied to these data the same analysis as described above. Some
examples of the 
 so-obtained \Dsabe\ curves, and their comparison with 
Eq.\eq{teor}, are shown in Fig.~\figsiete.  As it is evident in
this figure, the GGL approach with a total-energy cutoff
also explains these measurements, again in the
$10^{-2}\lsim\epsilon\lsim1$ range. These data also confirm the
existence of a well-defined reduced-temperature, \esuper, above
which the in-plane paraconductivity vanishes. We emphasize that in
performing these analyses we have again used \Tci\ as \Tc, and not
included any Maki-Thompson (MT) indirect contribution to the in-plane
paraconductivity (which, as commented in the Introduction, are now
well established to be absent in the
HTSC\cite{VeiraVidal,RamalloPRB}). As noted before,
Suzuki and Hikita assumed in their analyses the existence of such
MT contribution. In fact, to
be able to introduce the MT contribution, Suzuki
and Hikita had to consider the critical temperature as an
additional adjustable parameter. This leaded to \Tc's appreciably
different from \Tci.  Our present analysis of the data of Suzuki and Hikita also indicates that the in-plane paraconductivity in \LSCO\ films grown in (100)SrTiO$_3$ substrates is, as it could be expected, almost independent of thickness in the range \mbox{150 -- 350~nm}. However, let us stress again that using bulk samples, or films grown on different substrates or with significantly different thickness, may change some of the fine details of the paraconductivity, as in particular the  corresponding \xico\ values (see also below).

In  Fig.~\figcuatrob\ we  show the
comparison between the experimental \Dsabe\ measured in the present
work
 and the paraconductivity expressions without any cutoff and under the
conventional momentum or kinetic-energy cutoff, using the same values for
\xico\ and \esuper\ as in Fig.~\figcuatro. This figure evidences that, as it could
be expected,  the momentum or kinetic-energy cutoff expressions are able to fit
the experimental paraconductivity in a lower
$\epsilon$-range than the total-energy cutoff (up to at most
$\epsilon\simeq0.3$ for the momentum cutoff, and up to
about
$\epsilon\simeq0.1$ for no cutoff). In particular, they
fail to reproduce the rapid fall-off of the
fluctuations in the higher-$\epsilon$ region,
$\epsilon\gsim0.3$.

Let us now discuss the doping-dependence of the parameters \xico\
and \esuper\ which results from the above fits using the GGL
approach under a total-energy cutoff. In Fig.~\figocho\ we show such
\xico\ and \esuper\ values as a function of the doping. Let us
first remark the trend  for the variations of \xico\ with the doping
level
$x$: For $x\leq0.15$, \ie,  in the underdoped and optimally-doped range, \xico\ is
found to be  approximately equal to 0.9~\AA. This
value corresponds to a LD dimensional crossover parameter 
$\BLD\simeq7.5\times10^{-2}$. For $x=0.20$  it is
$\xico\simeq0.5$~\AA\ (and hence $\BLD\simeq2\times10^{-2}$), and for
$x\simeq0.25$ and above it is 
$\xico\sim0$~\AA\ (so $\BLD\sim0$ and the fluctuations are 2D in all
the studied $\epsilon$-range). This \xico-versus-$x$ trend is
consequent with our observation that the fluctuation conductivity is
the same for the underdoped and optimally-doped \LSCO\ films, and
increases with $x$ in the overdoped ones. Let us also note that in Refs.~\refcite{Loquet,Sato2,Bozovic} it has been argued that the substrate may induce in \LSCO\ films $c-$direction stress, what in turn may change the interlayer tunneling. Therefore,  \LSCO\ films grown on different substrates or with significant different film thickness, and also \LSCO\ bulk samples, may have \xico\ values different to the ones found in the present work. In this regard, we have  found  through
measurements of  the fluctuation-induced magnetization that the
superconducting fluctuations are bidimensional in  underdoped
{\it bulk}
\LSCO\ samples, in contrast with our
present results for thin films.\cite{Bulk2D}  It has been also found in Refs.~\refcite{overdopedbulkuno,overdopedbulkdos} that the superconducting bidimensionality is smeared or suppressed in the overdoped bulk samples of the \LSCO\ family. The comparison of these observations  with our present results suggests then that the substrate effects in \LSCO\ films do affect \xico, and therefore the interlayer tunneling of Cooper pairs.

Concerning the other free parameter in our $\Dsab(\epsilon)$ fits,
\epsilonsuper, we find that its value remains the same within the
experimental uncertainties for all the studied \LSCO\ films (see
Fig.~\figocho\ and Table~I). Such a value, which taking  into
account its experimental uncertainty (mainly  associated with the
background subtraction, see below) is bounded by
$0.4\lsim\epsilonsuper\lsim1.1$, is in fair agreement with the value
$\sim0.6$ that may be crudely estimated on the grounds of the
mean-field $\epsilon$-dependence of \xiabe\ and the BCS value of
$\xiabo/\xiabsubo$ in the clean limit  (see subsection~\ref{teorbackg} and
Ref.~\refcite{VidalEPL}). We note  that the 
validity of such a simple estimate has been also
confirmed, usually with a lower error band, by  the analysis of
the superconducting fluctuations in various optimally-doped HTSC
and various clean and moderately dirty LTSC  (see Ref.~\refcite{VidalEPL}
and references therein). We note also that our previous analyses of the fluctuation magnetization in a bulk underdoped \LSCO\ leaded to $\epsilonsuper\simeq0.6\pm0.2$. This is consistent with the fact that, as already mentioned in subsection~\ref{teorbackg} and further elaborated in Ref.~\refcite{VidalEPL}, the appearance of the reduced-temperature cutoff \epsilonsuper\ is expected to occur in all superconductors independently of, \eg,  their structural character or of the dimensionality of the superconducting fluctuations.

\subsection{On the influence of the background and
\Tc\ choices\label{secuncertainties}}

It is of relevance to discuss in detail the main sources of 
uncertainties in the above analyses of the experimental
paraconductivity in terms of the extended GGL approach.  Let us
first consider  the uncertainties
associated with the extraction of the normal-state background
contribution,
\roabBT. As already mentioned, we have checked that varying the
lower limit of the background fitting region from its  value in our
analyses,
$\TLB=4.5\Tc$, does not significantly change the obtained
$\Dsab(\epsilon)$ curves, provided that \TLB\  is kept above
$\sim3.5\Tc$ (otherwise it would force the paraconductivity  to
become negligible at progressively lower
reduced-temperatures, therefore affecting \Tsuper\ and \esuper). 
Adding a 3rd.~degree polynomial term to
the background functionality does not appreciably change the
paraconductivity curves that we have obtained. We have also checked 
that using a variable-range-hopping contribution, $\rho_1\exp(T_0/T)^{1/4}$, 
instead of the $b/T$ term (as proposed, \eg, by Leridon and coworkers\cite{Leridon} 
when analyzing the high-$\epsilon$ paraconductivity of \YBCO\ films) does not change 
the obtained backgrounds, within a maximum of 2\% variation.  The more
appreciable source of uncertainty in the background subtraction
comes from its
freedom to deviate $\pm20\%$   at
$10^{-2}\lsim\epsilon\lsim0.1$  from the
\Dsab\  obtained with the  background fitted closer to the
transition (fitted in the
$T$-region
$1.6\,\Tc\lsim T\lsim2.7\,\Tc$, \ie,
$0.5\lsim\epsilon\lsim1$). The
$\pm20\%$ figure was chosen after verifying that it is a good measure
of the variations of such ``low-$\epsilon$ background'' when its
fitting region is somewhat varied. While this uncertainty has
relatively low impact on the $\Dsab(\epsilon)$
curves for $10^{-2}\lsim\epsilon\lsim0.1$, its influence
becomes stronger  in the high reduced-temperature region 
($\epsilon>0.1$), affecting mainly the precise value of
\epsilonsuper\ (or, equivalently, of \Tsuper). In particular,  lower
backgrounds indeed  lead  to lower \epsilonsuper\ and \Tsuper.
However, let us emphasize that this uncertainty does not
affect the qualitative shape of the sharp fall-off of \Dsab\ at high
reduced-temperatures, but only the precise location of such a
fall-off. In Fig.~\figseis\ we illustrate the uncertainty of the
\Dsabe\ curves associated to the choice of the background. The
limits of the shaded areas  correspond to the lower and higher
backgrounds obtained for each sample. This figure clearly
illustrates that  the background
uncertainty does not affect the low-$\epsilon$ region in a dramatic
way, and therefore its impact on the value of \xico\ is also
moderate. Table~I summarizes the corresponding \esuper\ and \xico\
uncertainty for all the \LSCO\ films measured in this work.

It may be also useful to study the influence of the \Tc\ choice in
our analyses. First of all,
we note that the
${\rm d}\roab/{\rm d}T$ peaks around
\Tci\ present  about 1~K of half-widths at half-maximum of the peak
(see Fig.~\figtres\ and Table~I). Taking into account the relatively
low critical temperature of the
\LSCO\ family, such widths  correspond  to
rather high reduced-temperatures, of the order of
$\epsilon\sim5\times10^{-2}$. This means that varying \Tc\ inside
the upper half of the \Tci\ peak may considerably affect the
paraconductivity versus reduced-temperature curves in a sizable
$\epsilon$-region of our \Dsabe\  fits. Such
uncertainty is illustrated in Fig.~\figcinco\ for the cases 
$x=0.10$ and 0.25: In this figure, the same
\Dsab\ data as  in Fig.~\figcuatro\ is plotted against
$\epsilon\equiv\ln(T/\Tc)$ for
\Tc\ values in between \Tcim\ and \Tcip\ (the lower and upper
temperature boundaries of the \Tci\ peak at half-maximum). Also
plotted is Eq.\eq{teor}  in those cases where it is possible to find
\xico\ and
\epsilonsuper\ values producing a valid agreement with such
\Dsabe\ data curves. As it is clearly illustrated by the
figure, using \Tc's below \Tci\ makes  \Dsabe\  to
increase, and \Tc's above \Tci\  makes  \Dsabe\  to
decrease. In the
case $x=0.10$, when  lowering \Tc\ it is possible to fit
Eq.\eq{teor} up to \Tc\ values $\Tc=\Tci-0.6$K (=\Tcim+0.2K).
This is accomplished by lowering \xico, while the parameter 
\epsilonsuper\ remains almost unchanged from \esuper=0.8. In fact,
using $\Tc=\Tci-0.6$K leads to a fit with already $\xico=0$~\AA;
because of this,  with lower \Tc's  Eq.\eq{teor} can no longer account
for the data. In the case of the 
$x=0.25$ film, however, $\xico=0$~\AA\ already corresponds to
\Tc=\Tci\ and hence the \Dsabe\ curves cannot be successfully fitted
by Eq.\eq{teor} if using any \Tc\ below \Tci. Let us now discuss
what happens when using \Tc's above \Tci\ instead of below. In
such a case, the \Dsabe\ curves decrease. Such a decrease can be to
some extent  reproduced by Eq.\eq{teor}, for all the doping levels, 
by increasing
\xico\ and slightly changing also \epsilonsuper. However, as shown
in Fig.~\figcinco, the resulting fits are always of somewhat inferior
quality than when using \Tci. To sum up, using \Tci\ as \Tc\
seems to be the best choice because of the consistency when
taking into account the data of films with different doping
levels, and because it provides better-quality fits than \Tc's
taken above \Tci. However, we think that it is useful to bear in
mind the conclusion that  relatively small deviations from
\Tci\ will have an appreciable impact on the values of
\xico. In particular, the \xico\ values obtained for each \LSCO\
film using \Tcip\ instead of \Tci\ are 1.8~\AA\ for $x=0.10$,
2.0~\AA\ for $x=0.12$, 1.9~\AA\ for $x=0.15$, 1.6~\AA\ for
$x=0.20$, and 1.5~\AA\ for $x=0.25$. Also, for all the doping
levels studied here it is possible to find a \Tc\ in between
\Tcim\ and \Tci\ leading to $\xico\simeq0$~\AA. In fact,   such
uncertainties accumulate to the reasons commented in the
Introduction to explain the discrepancies between different authors
when proposing values for
\xico\ from the analysis of the superconducting fluctuations in
\LSCO. For instance, in our previous work \refcite{BriefReport}
we briefly analyzed a single
\LSCO\ film, with $x=0.10$, assuming $\xico\simeq0$~\AA,  what
leaded us to conclude that \Tc\ was nearer to \Tcim\ than  to \Tci. 
Finally, let us emphasize here that when using in our present
analyses  any choice of \Tc\ with similar displacements
with respect to \Tci\ for all the
$x$-values it remains true our main qualitative
conclusions: The paraconductivity is
constant with $x$ in the underdoped and optimally-doped \LSCO\ thin films,
but increases with $x$ in the overdoped range up to $x\simeq0.25$.
Also, the choice of
\Tc\ does not affect the fact that the fluctuations sharply decrease
at a well-defined reduced-temperature well above \Tc.

\section{Concluding remarks}

In this paper we have presented measurements of the in-plane
paraconductivity \Dsab\ above the superconducting transition of
various high-quality \LSCO\ thin films with thickness $\sim 150$~nm grown on (100)SrTiO$_3$
 substrates, with doping levels $x$ varying from 0.10 to 0.25, in
the reduced-temperature range $10^{-2}\lsim\epsilon\lsim1$. Our
results confirm, and extend to high reduced-temperatures, the
earlier proposals by Suzuki and Hikita\cite{Suzuki} and by Cooper
and coworkers\cite{Cooper} that in HTSC the variations of
\Dsab\ with doping may explain only a small part of the variations of
the total in-plane conductivity
\sab\ near
\Tc. In fact,  for the under- and optimally-doped compositions our
results directly show that the corresponding \roabT\ varies
appreciably whereas the
\Dsabe\  curves agree with each other well within the experimental
uncertainties. Our results also show  that in the
overdoped regime
\Dsabe\ increases only moderately with
$x$. The absence of important anomalous doping effects on the
paraconductivity is confirmed by the fact that our   \Dsabe\ data may
be appropriately  accounted for by the Gaussian-Ginzburg-Landau (GGL)
approach, extended to the high-$\epsilon$ region ($\epsilon\gsim0.1$)
by means of a total-energy cutoff. The fits using such an
approach lead to
$c$-direction superconducting coherence length amplitudes, \xico,  of about
0.9~\AA\ for the under- and optimally-doped films, about $0.5$~\AA\
for the $x=0.20$ overdoped film, and \xico\ negligibly small
($\sim0$~\AA) for the $x=0.25$ one. Moreover, independently of the
doping, we observe in all these
\LSCO\ films a rapid decrease of the superconducting
fluctuation effects in the high-$\epsilon$
region. Such a decrease is also well explained by the GGL
approach with a total-energy cutoff, which takes into
account the quantum localization energy associated with the
limits imposed by the uncertainty principle to the shrinkage of
the superconducting wave function  as
$\epsilon$ increases.\cite{VidalEPL}

In what concerns the fine behaviour of the in-plane paraconductivity
with doping, our results also confirm the
proposal by Suzuki and Hikita\cite{Suzuki} that in \LSCO\ films the
main effect of doping on the superconducting fluctuations is to
change the
$c$-direction superconducting coherence length, \xico. However,
because of the use of MT terms in their analyses, these
authors concluded that \xico\ would increase as $x$ increases in the
under-, optimally-, and over-doped regimes. In contrast, we have shown
here that when such terms are considered to be negligible one obtains
a different doping dependence of
\xico, as summarized above. As the absence of MT and DOS
contributions to the in-plane paraconductivity is at present well
established,\cite{VeiraVidal,RamalloPRB} we believe that the
true behaviour of \xico\ with doping  in \LSCO\ thin films grown in (100)SrTiO$_3$ substrates is probably the one we
propose here. 
Let us also note  that the dependence of \xico\ with doping found here  in \LSCO\ films may be not applicable to \LSCO\ bulk samples, as the substrate may also change \xico\ and, therefore, the dimensionality of the superconducting fluctuations. In particular, it has been  found in Ref.~\refcite{Bulk2D}, through
measurements of  the fluctuation-induced magnetization,  that the
superconducting fluctuations are bidimensional in  underdoped
{\it bulk}
\LSCO\ samples, in contrast with our
present results for thin films.  It has been also found in Refs.~\refcite{overdopedbulkuno,overdopedbulkdos} that the superconducting bidimensionality is smeared or suppressed in the overdoped bulk samples of the \LSCO\ family. These differences between thin films and bulk samples of \LSCO\ seem to be due  to strain effects associated
with the  grown on a substrate with $ab$-plane  lattice constants
 different to those of bulk
\LSCO.\cite{Chen,Loquet,Sato2,Bozovic}
Proposals that the main effect of doping on the superconducting
fluctuations is to change \xico\ were also done (and almost
simultaneously to Suzuki and Hikita\cite{Suzuki}) for the \YBCO\ compound by Cooper \etal\cite{Cooper}
(and later also by Juang \etal\cite{Juang}). We note however
that our present results cannot be directly extrapolated to 
\YBCO\ because in this compound doping happens in the interlayer
CuO-chains, absent in \LSCO.  
Such CuO-chains are expected to determine most of the interlayer
tunneling in 
\YBCO\ due to their metallic-like character.   Another central result of our analyses is that  the cutoff parameter, \esuper, arising in the ``extended'' GGL approach is found to be, well within the experimental uncertainties,
independent of the doping. As the conventional momentum cutoff is
just a particular case, valid for $\epsilon\lsim0.2$, of the
total-energy cutoff approach, our conclusion will then apply to any
analysis in terms of such a momentum cutoff. This suggests,
therefore, that the doping dependence of the momentum cutoff
observed in other HTSC compounds (see, \eg, Refs.~\refcite{Asaka} and
\refcite{Silva}) could be an extrinsic effect due to the presence of
sample inhomogeneities or to ambiguities in the background
estimate.

Our results may also have  implications on other
open problems of the HTSC. First of all, they allow us to conclude,
as mentioned in the title of this paper, that   the superconducting
fluctuations in underdoped \LSCO\
films are not linked to the opening of their normal-state
pseudogap: First, because the variations of doping in such
underdoped compounds, which are know to vary the pseudogap opening
temperature,
$T^*$,\cite{ReviewInicial}  do not result in any observable
change of the superconducting fluctuations, even in the
high-$\epsilon$ region. Second, because also in such underdoped
compositions  the superconducting fluctuations have been found here
to disappear at reduced-temperatures which are, even taken into
account the experimental uncertainties,  always below 1.1,
\ie, much below $T^*$ (that is located  above room
temperature for the underdoped compositions studied
in this work\cite{ReviewInicial}). In other words, our
experimental results seem to be contradictory with  the theoretical
proposals that in HTSC the superconducting fluctuations are strongly
enhanced when
underdoping.\cite{ReviewInicial,Meingast,Varlamov,Randeria,Emery2D,EmeryStripes}
Such proposals include, \eg,  the ones   that   the pseudogap effects observed in
the underdoped HTSC are due to strong fluctuations of the superconducting order
parameter (and in particular of its phase), which in  turn would be due to
either a Bose-Einstein--like preformation of Cooper pairs\cite{Randeria},
stripe-like inhomogeneities\cite{Varlamov,EmeryStripes}, or bidimensionality
effects\cite{Emery2D}. For instance, in Ref.~\refcite{Randeria} it is
claimed that the superconducting fluctuations in underdoped HTSC will
be given essentially by the 2D-Kosterlitz-Thouless model of
fluctuations\cite{Nelson} with relaxation times of such
fluctuations various orders of magnitude bigger than the BCS mean-field one
implied by Eq.\eq{teor}. However, our present analyses strongly
suggest that there is no change in the order of magnitude of the
superconducting fluctuations when the doping level enters the
underdoped range. There is also no evidence in our experimental
data of any dependence on $x$ of the critical exponents of
\Dsabe\ for $0.10\leq x\leq0.15$. Our results favor then the
theoretical proposals associating the pseudogap opening to purely
normal-state effects (as, \eg, those of Refs.~\refcite{Pines,Bok,Anderson}), or the
theories   where the Bose-Einstein preformed pairs are dressed by
normal quasiparticles\cite{Larkin}. In those approaches, 
consistently with our present results,
the superconducting
order parameter is expected to undergo Gaussian, mean-field--like
fluctuations, except in the very close vicinity of \Tc\
(for $\epsilon\lsim10^{-2}$).  It would be useful to determine
whether other proposals for the doping effects in HTSC, such as
the existence of a $T=0$~K quantum transition near optimal
doping,\cite{Griego} would lead to Gaussian or non-Gaussian
superconducting fluctuations above \Tc.

The  results presented here may also have  implications on  the theoretical
proposals by Horbach and coworkers\cite{DSMFL} for the
paraconductivity in the marginal-Fermi-liquid (MFL) scenario, which
would apply  to the   optimally-doped and overdoped HTSC. According
to the calculations in
Ref.~\refcite{DSMFL}, the inelastic  scattering of normal quasiparticles
in optimally-doped and overdoped HTSC would decrease the
relaxation time   of the superconducting
fluctuations below the BCS mean-field value, and correspondingly
also
\Dsab. Horbach and coworkers estimate such changes in \Dsab\ to be of
the order of a prefactor 0.2-0.75 (depending mainly on a cut
parameter affecting their numerical evaluations).\cite{DSMFL} However, our
present results indicate that  the  relaxation
time of the superconducting fluctuations takes  the BCS mean-field
value for all the doping levels. They also indicate that \Dsab\ is not smaller in
the optimally-doped and overdoped films than it is in the underdoped films [in
fact, it is even higher in the overdoped films because of their lower \xico].

It may be also useful  to compare our  results for the
\xico\ dependence on doping with detailed measurements of the
normal-state anisotropy in \LSCO\ films. For instance, in
Ref.~\refcite{inclinados} it is measured the resistivity in \LSCO\
thin films grown with the
$c$-axis oriented obliquely with respect to the substrate. From such
measurements, it was proposed that the normal-state anisotropy
decreases as doping increases. If confirmed by different
measurements, this  would indicate that the interlayer
tunneling of the normal  and superconducting carriers  would not be
directly related. This could be coherent, \eg, with the interlayer
tunneling model for the
 superconducting condensation in HTSC proposed 
 by Anderson and coworkers.\cite{Anderson} Further work to compare
the normal and superconducting interlayer tunnelings in \LSCO\ films
is clearly needed, both  theoretical and experimental (for instance,
through measurements of the intrinsic interlayer Josephson effects
and of the normal and superconducting magnetoconductivity). 
Further work is also needed to study the possible influence on the {\it ``fine''} details of  the
superconducting fluctuations of the film thickness and of the type of substrate. 
As mentioned previously, both factors are known to vary  several  crucial properties of the \LSCO\
films, like their
\Tc\ and
\roab(250~K) values, possibly due in part to strain effects associated
with the  grown on a substrate with $ab$-plane  lattice constants
 different to those of bulk
\LSCO.\cite{ReviewInicial,Suzuki,Kimura,Chen,Loquet,Sato2,Bozovic,Kao} In the present paper we have found a transversal superconducting coherence length for \LSCO\ films which is different, both in value and doping dependence, to the one that seems to arise from Ref.~\refcite{Bulk2D} (where the fluctuation magnetization in a bulk underdoped \LSCO\ was measured) and  Refs.~\refcite{overdopedbulkuno,overdopedbulkdos} (where the superconducting anisotropy of \LSCO\ bulk samples with different dopings was measured). It would be interesting,
therefore, to further study the relationships between sample
thickness, type of substrate, and superconducting interlayer
tunnelings  [or, equivalently,
\xico].

\mbox{}

\nonfrenchspacing
This work has been supported by the CICYT, Spain, under grants
no.~MAT2001-3272 and MAT2001-3053, by the Xunta de Galicia under
grant PGIDT01PXI20609PR, and by Uni\'on Fenosa under grant 0666-2002.

\mbox{}

\newpage

%%%%----TABLE-1-----%%%%

\mbox{}\vspace{3cm}\mbox{}\\

\epsfxsize=\textwidth
\centerline{\epsfbox{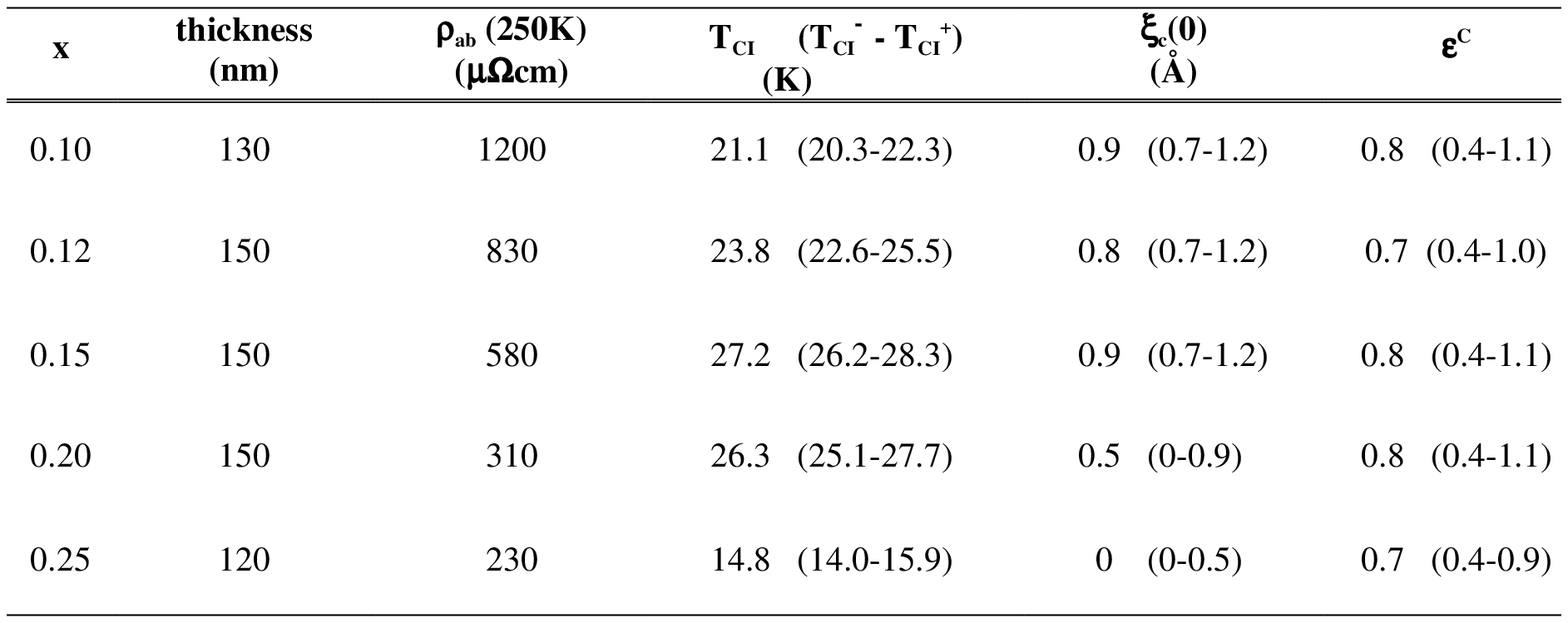}}

\mbox{}

{Table~I:\\ \mbox{}\\ \footnotesize Summary of the main parameters of the \LSCO\ thin films
studied in this work. The \Tci, \Tcim\ and \Tcip\ temperatures
are represented in Fig.~\figtres. The
\xico\ and
\epsilonsuper\ main values are the ones  resulting from the
\Dsabe\ fits presented in Fig.~\figcuatro. Their uncertainties  result
from considering different normal-state backgrounds  (see
Fig.~\figseis\ and main text).}

\newpage

%%%%%%%%%---Fig.1---%%%%%%%%
\begin{figure}[ht]
%\mbox{}\vspace{-1cm}\mbox{}\\
\epsfxsize=0.42\textwidth
\centerline{\epsfbox{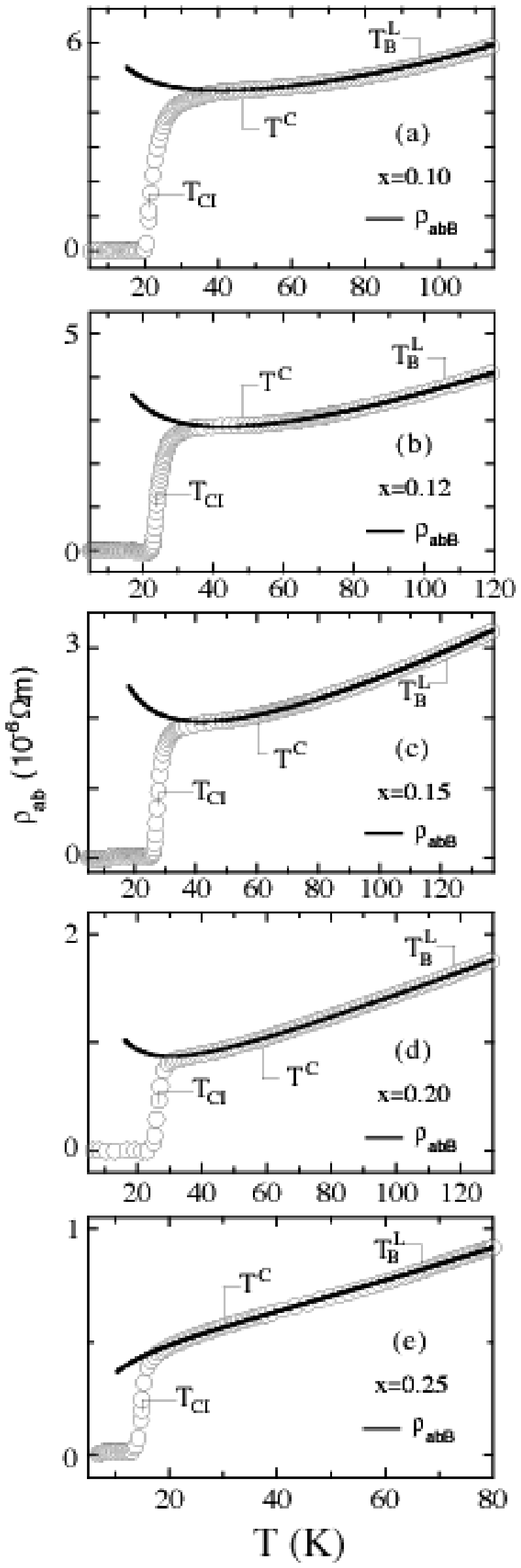}}
%\mbox{}\vspace{-2.5cm}\mbox{}\\
Figure 1:
\footnotesize \protect\mbox{}\protect\\ \protect\mbox{}\protect\\ 
The in-plane resistivity measured in this work
(circles) in various \LSCO\ thin films with different doping levels
$x$. We also indicate (solid line) the normal-state background of each
film, obtained by using the procedure indicated in the main
text. The temperature 
\TLB\ corresponds to the lower limit of the fitting region used to
extract such background. The temperature
\Tsuper\ is the one at which the background and the
experimental data  first deviate from each other (\ie, the
temperature above which fluctuation effects are no longer
observed). Note that \Tsuper\ is located well below \TLB\ (see also
main text for details).  \Tci\ is the temperature where ${\rm d}\roab/{\rm d}T$ is maximum, that is expected to be close to \Tc, the mean-field normal-superconductor transition temperature.
\end{figure}

\newpage

%%%%%%%%%---Fig.2---%%%%%%%%

\begin{figure}[ht]
%\mbox{}\vspace{-1cm}\mbox{}\\
\epsfxsize=0.35\textwidth
\centerline{\epsfbox{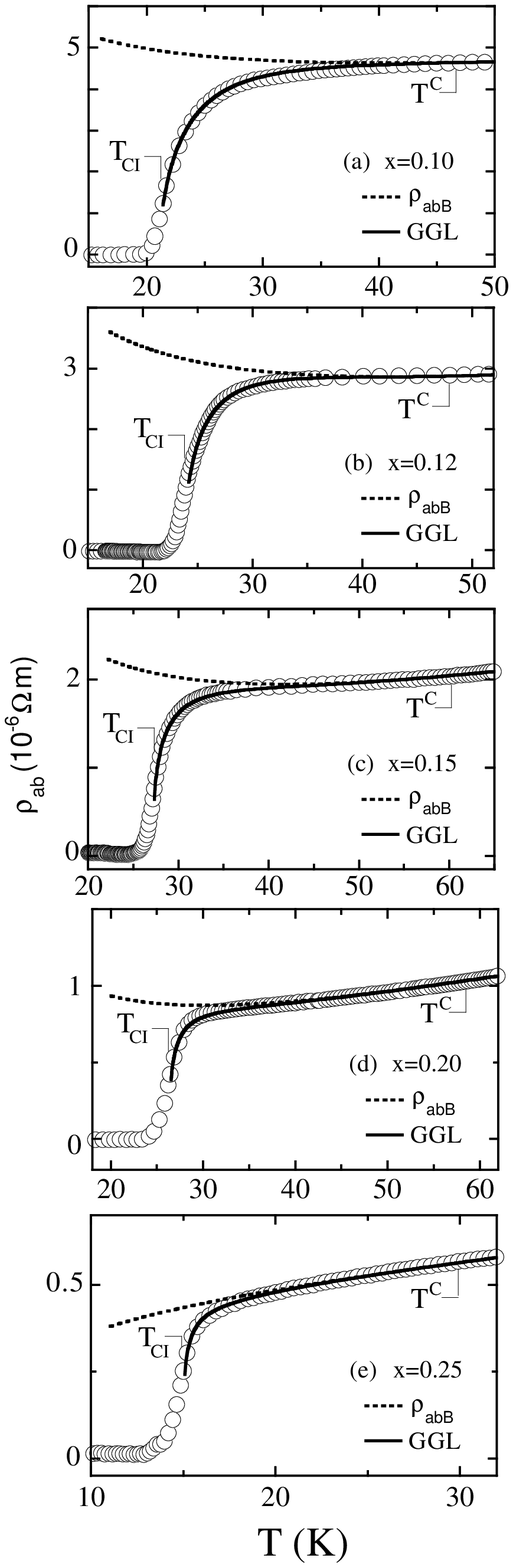}}
%\mbox{}\vspace{-2.5cm}\mbox{}\\
Figure 2:
\footnotesize \protect\mbox{}\protect\\ \protect\mbox{}\protect\\ 
   Amplification of the view in Fig.~\figuno\ to better
appreciate the range of temperatures where the fluctuation
effects above the transition are observed. The dashed line
corresponds to the normal-state background and the solid line
corresponds to the theoretical \roabT\ obtained  by
using in Eq.~\eq{contribuciones} such a background and the best fit
by the GGL theory extended to high reduced-temperatures through its regularization with a so-called ``total-energy'' cutoff which takes into account the quantum localization energy.
\end{figure}

\newpage

%%%%%%%%%---Fig.3---%%%%%%%%

\begin{figure}[ht]
%\mbox{}\vspace{-2.5cm}\mbox{}\\
\epsfxsize=0.35\textwidth
\centerline{\epsfbox{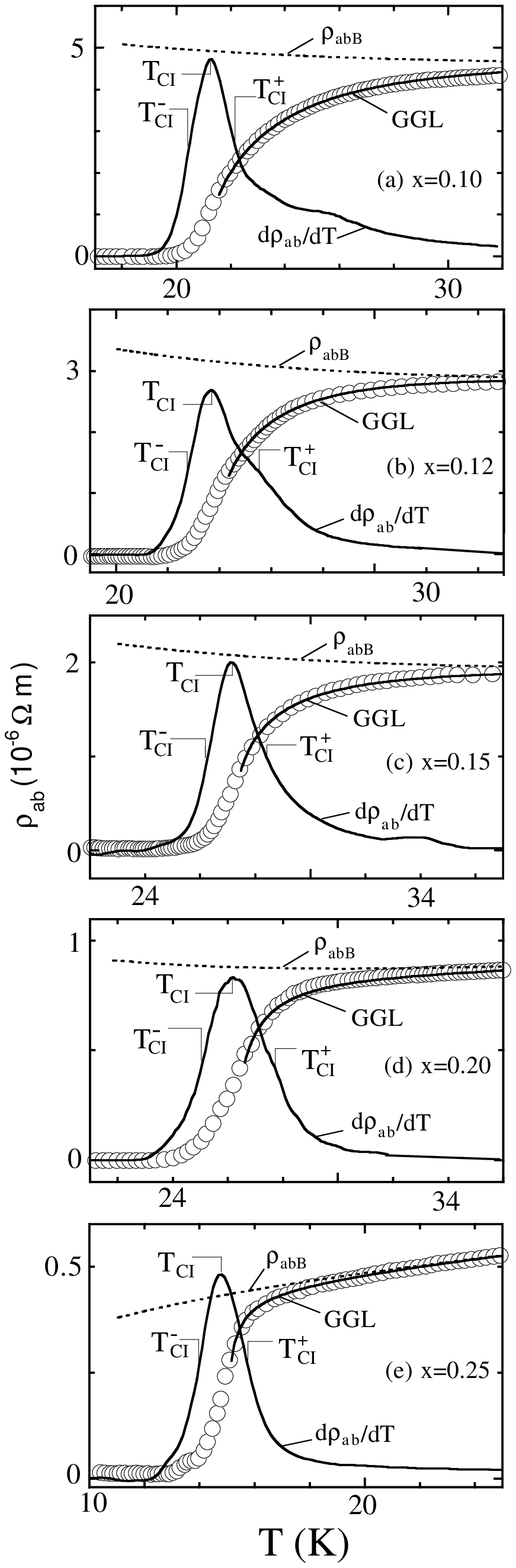}}
%\mbox{}\vspace{-2.5cm}\mbox{}\\
Figure 3:
\footnotesize \protect\mbox{}\protect\\ \protect\mbox{}\protect\\ 
 Scoop of Figs.~\figuno\ and \figdos\ to better
appreciate the region closer to \Tci,  the temperature where ${\rm
d}\roab/{\rm d}T$ is maximum. Such a ${\rm d}\roab/{\rm
d}T$ peak is also plotted, in arbitrary units (as a solid 
line). The temperatures \Tcim\ and \Tcip\ correspond to the
half-maximum boundaries of such a peak.  The circles are the
experimental \roabT\ data, the dashed line is the normal-state
background, and the solid line following the data is the  \roabT\
resulting from such a background and the  ``extended'' GGL
theory with a total-energy cutoff. 
\end{figure}

\newpage

%%%%%%%%%---Fig.4---%%%%%%%%

\begin{figure}[ht]
\epsfxsize=0.55\textwidth
\centerline{\epsfbox{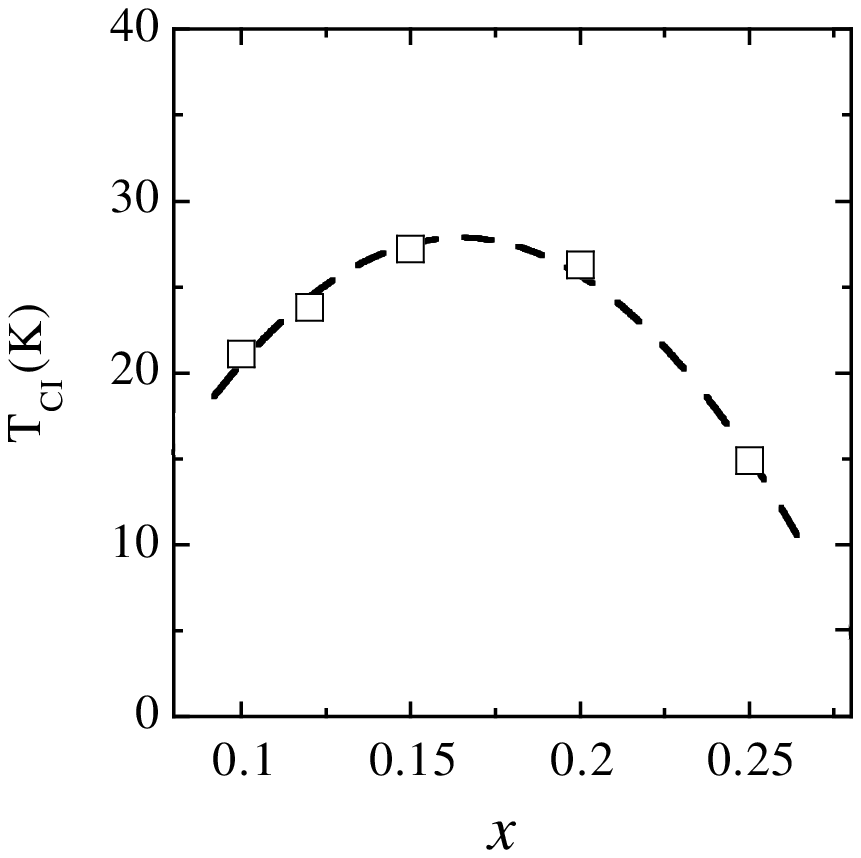}}
%\mbox{}\vspace{-2.5cm}\mbox{}\\
Figure 4:
\footnotesize \protect\mbox{}\protect\\ \protect\mbox{}\protect\\ 
 The temperature \Tci\ where it is located the
inflexion point of the \roabT\ transition (see Figs.~\figdos\ and
\figtres) versus the Sr-content $x$ in the \LSCO\ films measured in
this work. The dashed line is a parabolic fit which serves as a
guide for the eyes. Note that both the absolute values and the
doping dependence of such \Tci\ is in fairly good agreement with
the one well 
established\cite{ReviewInicial,Suzuki,Loquet,Sato2,Kao,BriefReport}
for the critical temperature of \LSCO\ films grown on  (100)SrTiO$_3$ substrates and
with similar thickness. 
\end{figure}

\newpage

%%%%%%%%%---Fig.5---%%%%%%%%

\begin{figure}[ht]
\mbox{}\vspace{0.5cm}\mbox{}\\
\epsfxsize=0.50\textwidth
\centerline{\epsfbox{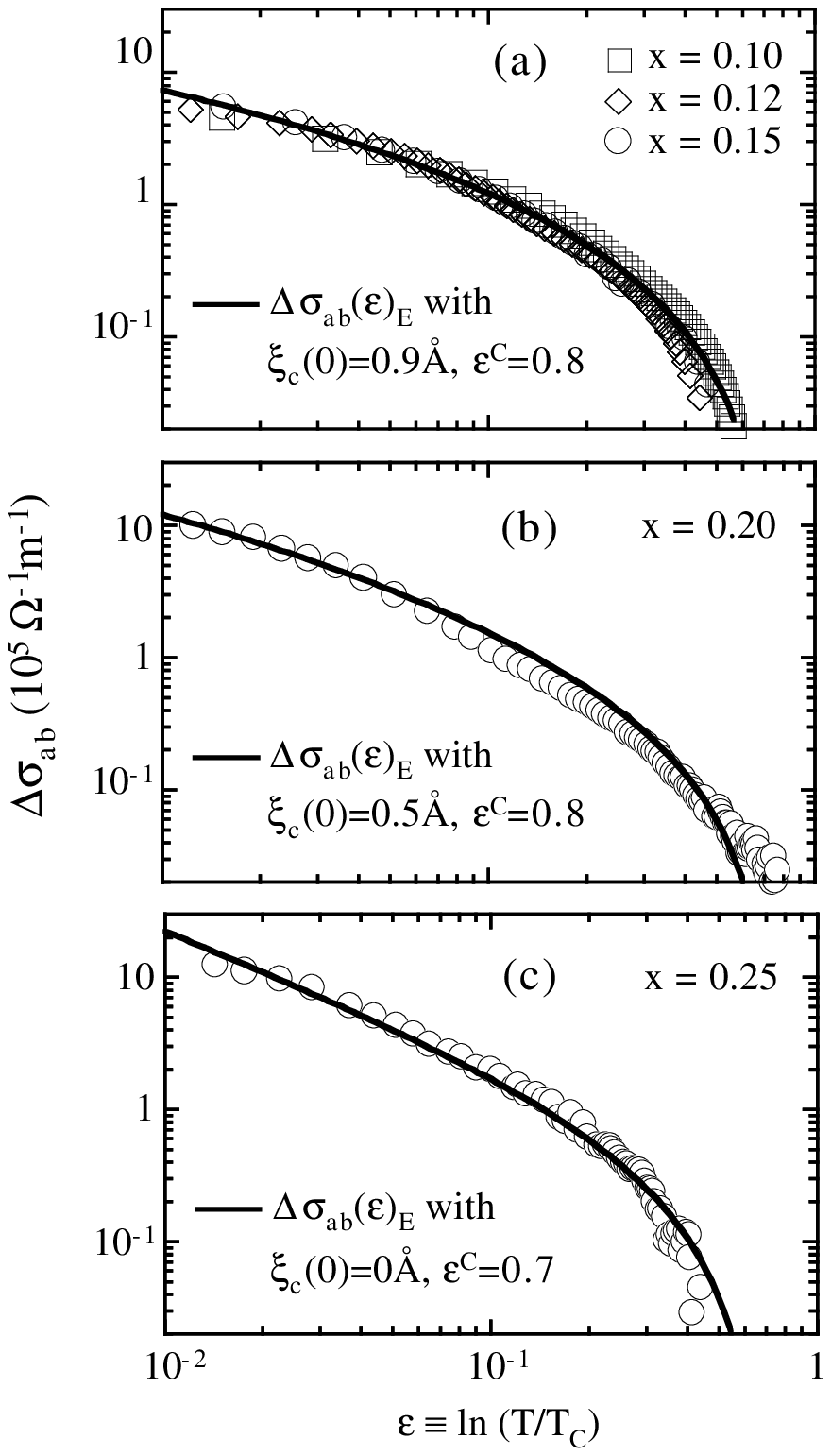}}
%\mbox{}\vspace{-2.5cm}\mbox{}\\
Figure 5:
\footnotesize \protect\mbox{}\protect\\ \protect\mbox{}\protect\\ 
 Comparison between the ``extended'' GGL expressions 
for \Dsabe\
using a total-energy cutoff [solid lines, Eq.\eq{teor}] and the
experimental curves of the in-plane paraconductivity  versus 
reduced-temperature in the \LSCO\ films measured in this work
(circles). In these comparisons we use
\Tci\ as critical temperature (see main text). As it can be seen in
Fig.~(a), the experimental \Dsabe\ curves for the underdoped and
optimally-doped \LSCO\ films are essentially coincident. The
corresponding \Dsabe\  for the overdoped films [Figs.~(b)
and (c)] show a moderate increase as $x$ increases. The GGL
fits are able to account for such data  in all the studied
$\epsilon$-range, including also the disappearance of observable
fluctuation effects at high reduced-temperatures. See main text for
details. 
\end{figure}

\newpage

%%%%%%%%%---Fig.6---%%%%%%%%

\begin{figure}[ht]
\mbox{}\vspace{2cm}\mbox{}\\
\epsfxsize=0.60\textwidth
\centerline{\epsfbox{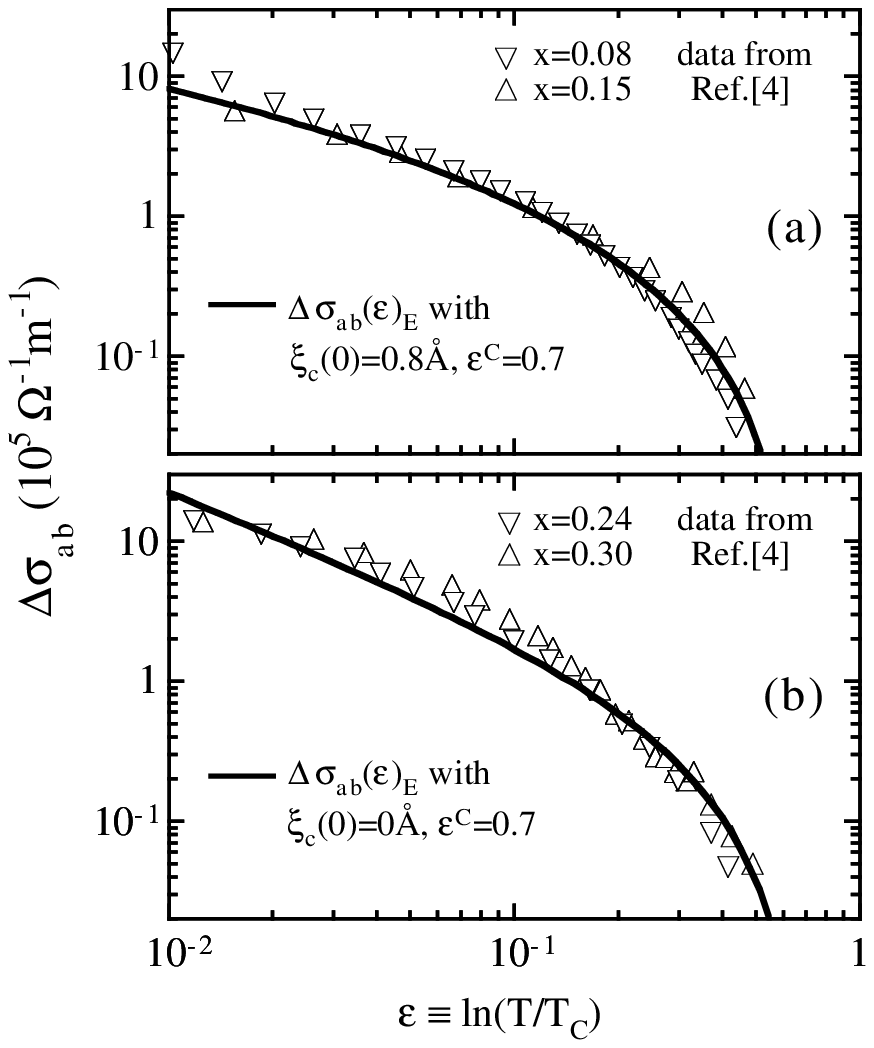}}
%\mbox{}\vspace{-2.5cm}\mbox{}\\
Figure 6:
\footnotesize \protect\mbox{}\protect\\ \protect\mbox{}\protect\\ 
 Some examples of the in-plane paraconductivity
versus reduced-temperature curves that result from
applying to the \roabT\ data measured  by Suzuki and
Hikita\cite{Suzuki} in \LSCO\ films  the same analyses as we have applied
to our present measurements. The solid lines correspond to the fits using
the ``extended'' GGL approach with a total-energy cutoff. Such fits fully confirm
the results  found with our own measurements (compare, \eg, with
Fig.~\figcuatro). 
\end{figure}

\newpage

%%%%%%%%%---Fig.7---%%%%%%%%

\begin{figure}[ht]
\mbox{}\vspace{0.0cm}\mbox{}\\
\epsfxsize=0.50\textwidth
\centerline{\epsfbox{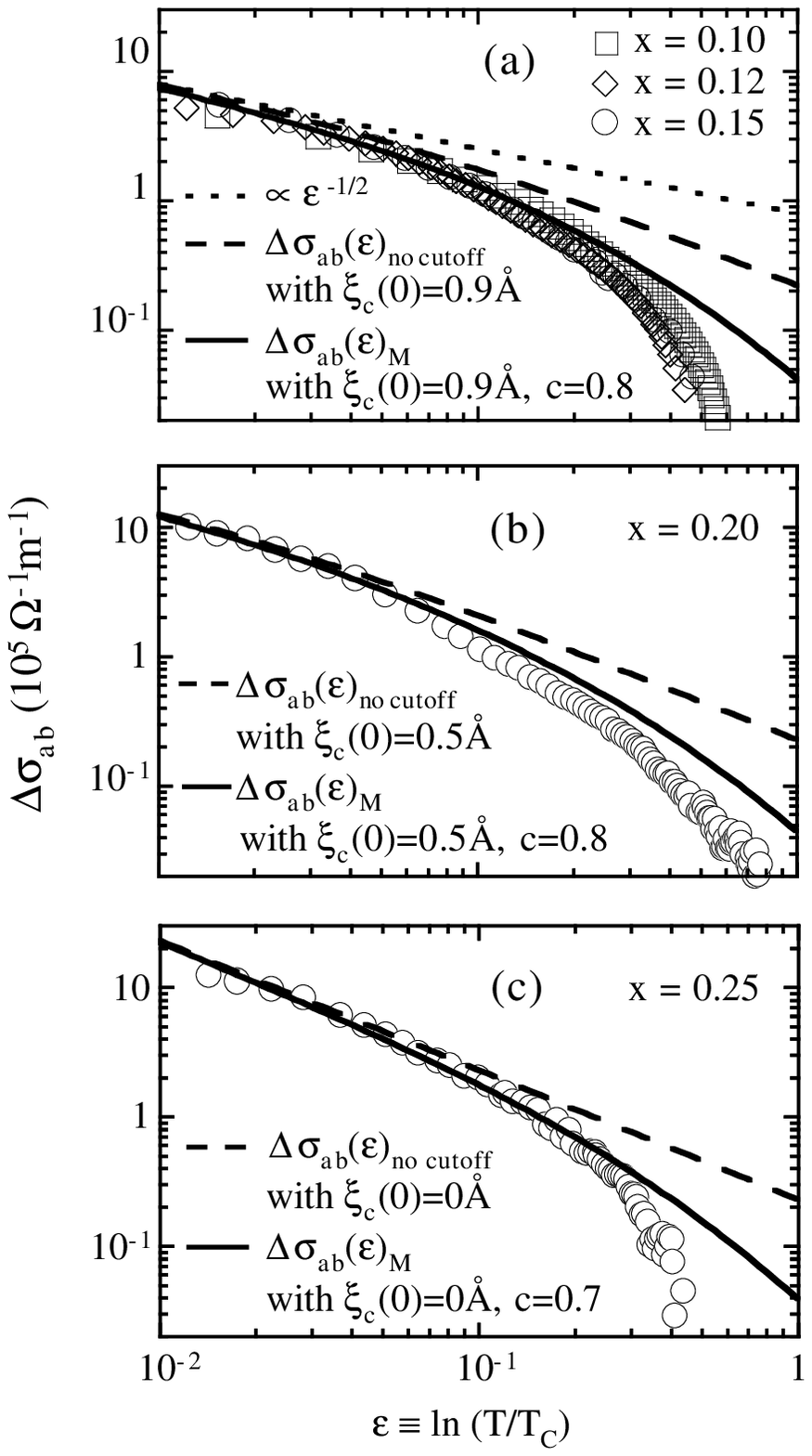}}
%\mbox{}\vspace{-2.5cm}\mbox{}\\
Figure 7:
\footnotesize \protect\mbox{}\protect\\ \protect\mbox{}\protect\\ 
 Comparison between the same experimental curves as
in Fig.~\figcuatro\ and the paraconductivity expressions without any
cutoff [dashed lines, Eq.\eq{teorLD}] and with the momentum or
kinetic-energy cutoff [continuous lines, Eq.\eq{teorM}]. In
performing these comparisons, we have used the same values for
\xico\ and the cutoff constant as in Fig.~\figcuatro. In
Fig.~\figcuatrob (a) we also indicate the low-$\epsilon$  asymptotic
$\Dsabe\propto\epsilon^{-1/2}$, which would correspond to the 3D
limit without any cutoff. The dashed line in Fig.~\figcuatrob (c)
corresponds also to $\Dsabe\propto\epsilon^{-1}$, \ie, the 2D
limit without any cutoff. The $\epsilon\sim10^{-2}$
paraconductivity of the 
$x=0.20$ film does not follow any of such dimensionality
 limit cases, but lies instead in the dimensional crossover
regime. Note that the momentum or kinetic-energy cutoff
expressions are able to explain the experimental paraconductivity
in a lower
$\epsilon$-range than the total-energy cutoff (up to at most 
$\epsilon\simeq0.3$).
The expressions without any cutoff explain the data in a even more limited
$\epsilon$-range, 
$\epsilon\lsim0.1$. 
\end{figure}

\newpage

%%%%%%%%%---Fig.8---%%%%%%%%

\begin{figure}[ht]
\mbox{}\vspace{1cm}\mbox{}\\
\epsfxsize=0.50\textwidth
\centerline{\epsfbox{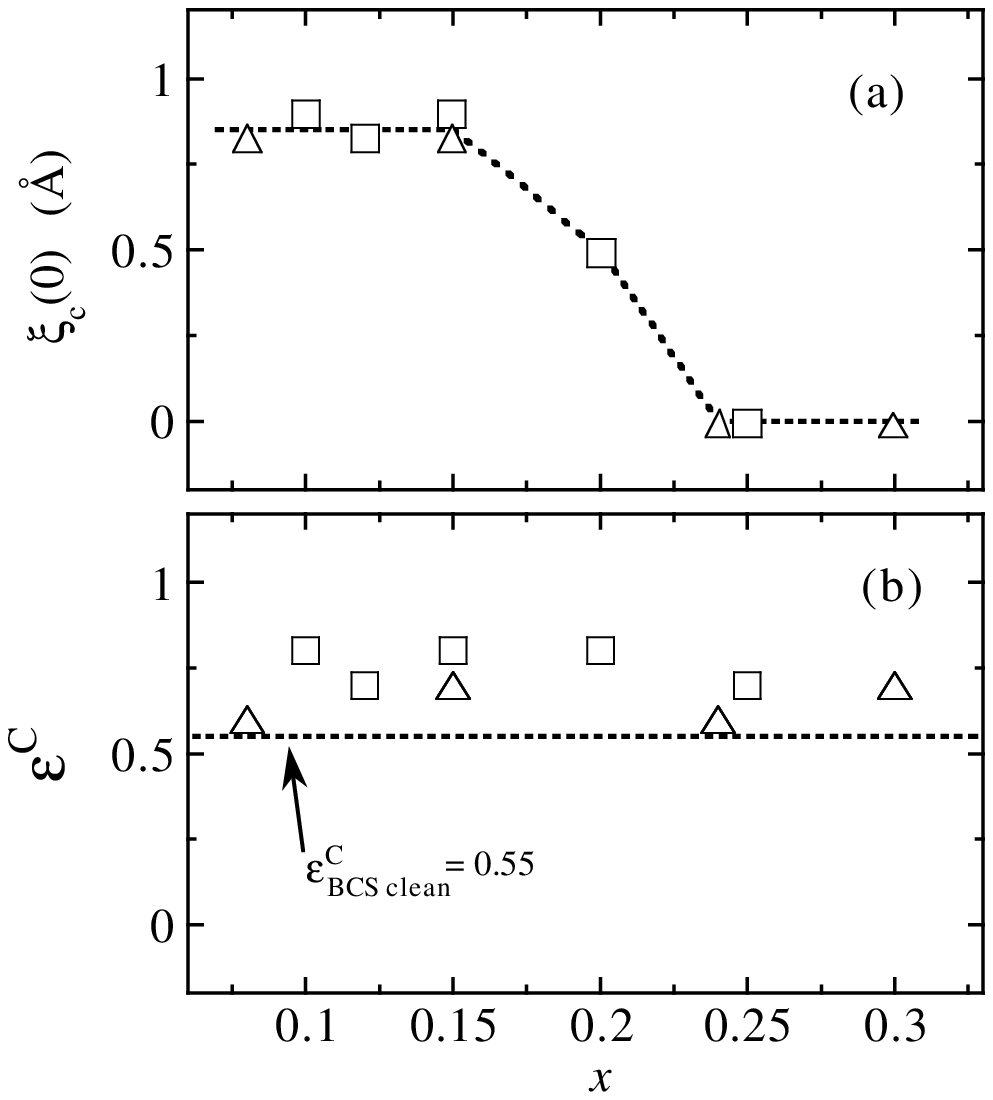}}
%\mbox{}\vspace{-2.5cm}\mbox{}\\
Figure 8:
\footnotesize \protect\mbox{}\protect\\ \protect\mbox{}\protect\\ 
 The values of the $c$-direction coherence length
GL amplitude,
\xico, and of the total-energy cutoff amplitude, \esuper, 
providing the best fit to the \Dsabe\ data (see Figs.~\figcuatro\ and
\figsiete), represented versus the Sr-content (\ie, hole doping) $x$.
The squares correspond to the \LSCO\ films measured in the present
work, while the triangles correspond to the films measured by
Suzuki and Hikita\cite{Suzuki} analyzed in Fig.~\figsiete. As
it can be easily seen in Fig.(a), 
\xico\ is constant, well within the experimental
uncertainties, for the underdoped ($x<0.15$) and optimally-doped
($x=0.15$) films. When entering in the overdoped range ($x>0.15$),
\xico\ decreases up to being negligible [$\xico\simeq 0$~\AA] at about
$x\simeq0.25$ and above. The dashed line in this Fig.(a) is a guide
for the eyes.  Figure (b) illustrates that the cutoff
amplitude \esuper\  is found to be, within the
experimental uncertainties, constant for all doping levels. The
continuous line in this Fig.(b) is the crude estimate
$\epsilon^C_{\rm BCS clean}\simeq0.6$ that can be obtained  by using
the BCS theory in the clean limit (see main text). Note that the
\esuper\ values shown in this figure are somewhat above such an
estimate. However, such differences cannot be taken as significative,
in view of the error bars of 
\esuper\ in our \Dsabe\ analyses (see main text and Table~I). 
\end{figure}

\newpage

%%%%%%%%%---Fig.9---%%%%%%%%

\begin{figure}[ht]
\mbox{}\vspace{1cm}\mbox{}\\
\epsfxsize=0.60\textwidth
\centerline{\epsfbox{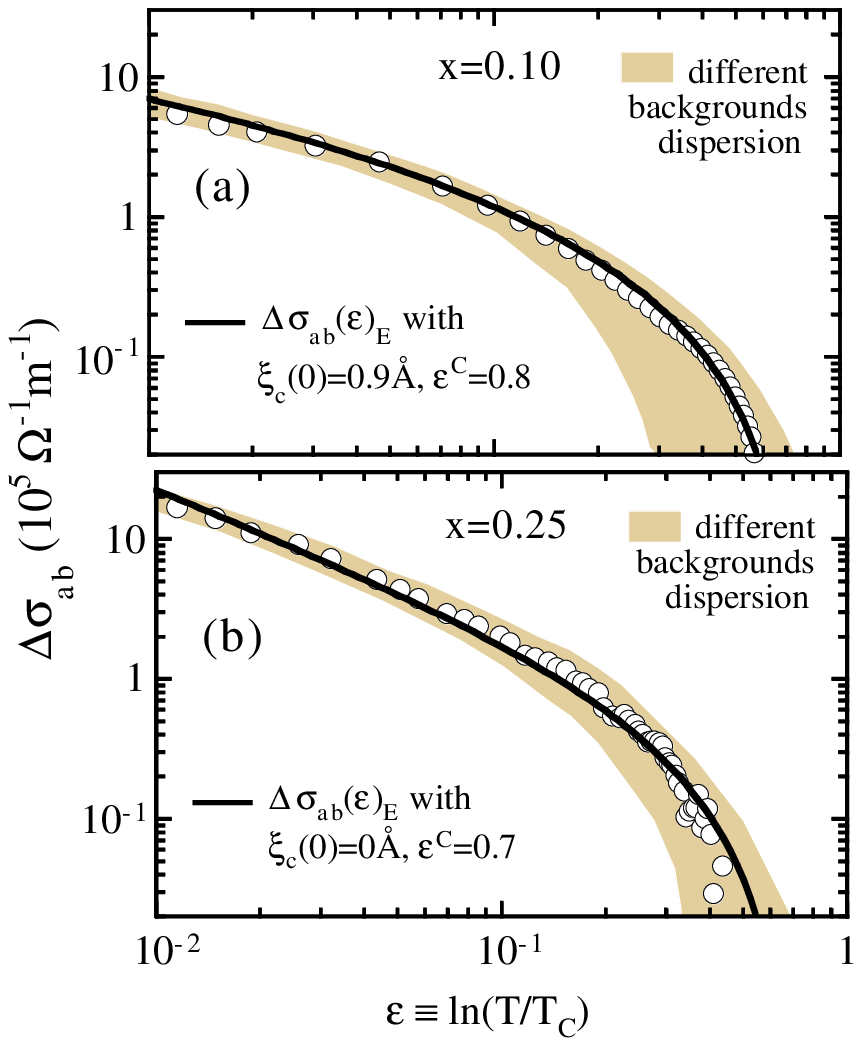}}
%\mbox{}\vspace{-2.5cm}\mbox{}\\
Figure 9:
\footnotesize \protect\mbox{}\protect\\ \protect\mbox{}\protect\\ 
 Influence of the use of different normal-state backgrounds  in
the experimental
\Dsabe\ curves of  (a) the $x=0.10$ film and (b) the
$x=0.25$ film measured in this work.  Such a background uncertainty is discussed in
detail in the main text. The corresponding uncertainties in
\esuper\ and \xico\ are summarized in 
Table~I for all the films measured in this work. 
\end{figure}

\newpage

%%%%%%%%%---Fig.10---%%%%%%%%

\begin{figure}[ht]
\mbox{}\vspace{1cm}\mbox{}\\
\epsfxsize=0.55\textwidth
\centerline{\epsfbox{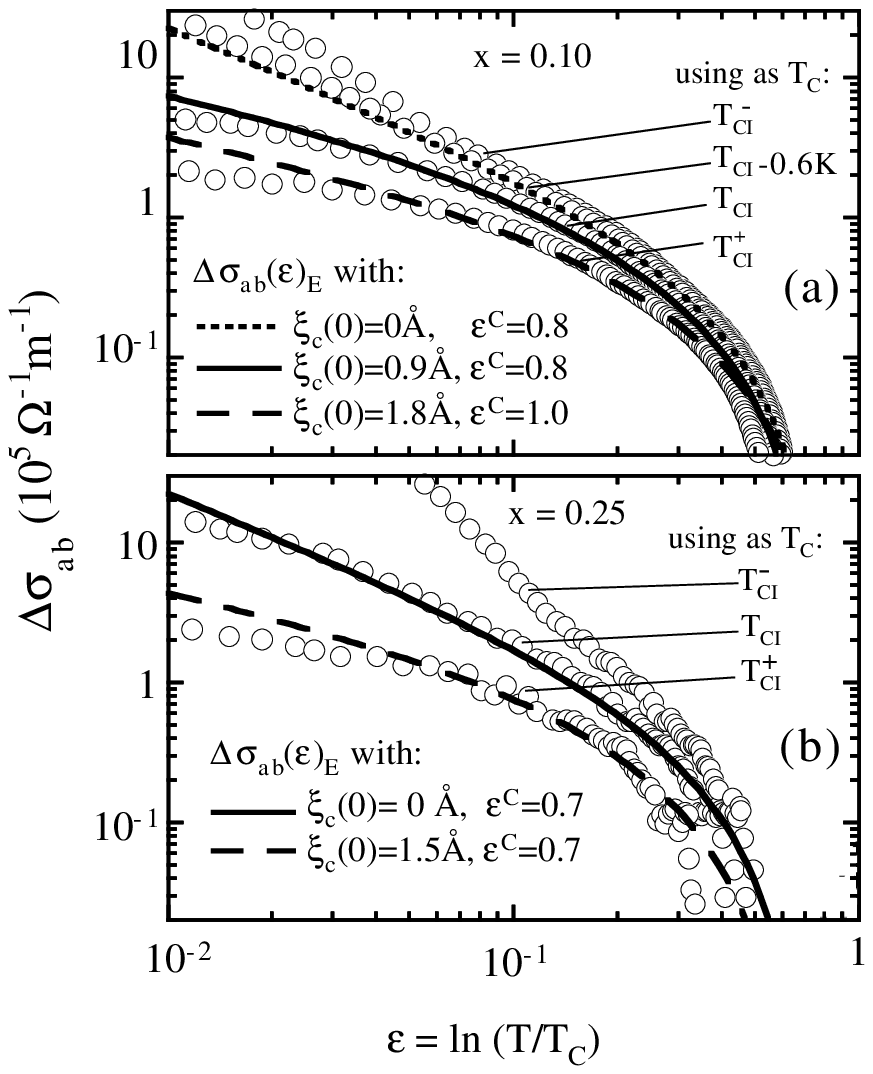}}
%\mbox{}\vspace{-2.5cm}\mbox{}\\
Figure 10:
\footnotesize \protect\mbox{}\protect\\ \protect\mbox{}\protect\\ 
 Influence of the choice of \Tc\ in the experimental
\Dsabe\ curves and GGL fits, for  (a) the $x=0.10$ film and (b) the
$x=0.25$ film measured in this work. 
The 
\Tc\ choice varies mainly the
\Dsabe\ slope in the
$\epsilon\lsim0.1$ region.
As easily visible in these figures, using \Tc's below \Tci\ does
not allow to fit the data with Eq.\eq{teor} for all the doping
levels, and \Tc's above \Tci\ produce poorer fits than \Tci.
The other underdoped
and optimally-doped films studied in this work produce curves
similar to the $x=0.10$ ones, and the
$x=0.20$ film produces results intermediate between those of (a) and
(b).  
\end{figure}

\end{document}